\documentclass[prb,showpacs,preprintnumbers,preprint]{revtex4}
\usepackage{graphicx}
\begin{document}
\def\rhov{{\mbox{\boldmath{$\rho$}}}}
\def\tauv{{\mbox{\boldmath{$\tau$}}}}
\def\Lambdav{{\mbox{\boldmath{$\Lambda$}}}}
\def\lambdav{{\mbox{\boldmath{$\lambda$}}}}
\def\sigmav{{\mbox{\boldmath{$\sigma$}}}}
\def\xiv{{\mbox{\boldmath{$\xi$}}}}
\def\chiv{{\mbox{\boldmath{$\chi$}}}}
\def\rhov{{\mbox{\boldmath{$\rho$}}}}
\def\deltav{{\mbox{\boldmath{$\delta$}}}}
\def\phiv{{\mbox{\boldmath{$\phi$}}}}
\def\piv{{\mbox{\boldmath{$\pi$}}}}
\def\psiv{{\mbox{\boldmath{$\psi$}}}}
\def\Psiv{{\mbox{\boldmath{$\Psi$}}}}
\def\oh{{\scriptsize 1 \over \scriptsize 2}}
\def\ot{{\scriptsize 1 \over \scriptsize 3}}
\def\tt{{\scriptsize 3 \over \scriptsize 2}}
\def\of{{\scriptsize 1 \over \scriptsize 4}}
\def\tf{{\scriptsize 3 \over \scriptsize 4}}
\title{Landau Potentials for Multiferroic Mn$_2$GeO$_4$}

\author{A. B. Harris}

\affiliation{ Department of Physics and Astronomy,
University of Pennsylvania, Philadelphia, PA 19104}
\date{\today}

\begin{abstract}
It is shown how to adapt the results of ISODISTORT to a more convenient
form.  For the case of Mn$_2$GeO$_4$ we characterize the complex magnetic
phase that exists at tmperature below 5.5K, by order parameters for
both the commensurate ordering and the incommensurate ordering. For
the incommensurate ordering we are forced to consider the transformation
properties which interrelate magnetic modes at different noncollinear members
of the star of the incommensurate wave vector.  The transformation
properties of the order parameters and of the underlying magnetic
wave functions are developed. These results are applied to construct
the high order invariants in the free energy which have been used elsewhere to
describe the characteristics of switching between different
domains.  
\end{abstract}
\maketitle


\section{INTRODUCTION}

The last dozen years have seen an explosion in the study of multiferroic
systems in which incommensurate magnetic order induces 
ferroelectricity, following the pioneering work of
Refs. \onlinecite{KIMURA}, \onlinecite{HUR}, and \onlinecite{GOTO}.
Shortly thereafter a microscopic model was developed\cite{CURRENT}
based on the idea of a ``spin current.''   This mechanism has been
widely cited in terms of a picture in which the spin structure is
characterized as being a magnetic spiral.\cite{MOSTOVOY}
However, this idealized picture is not always easy to identify
when the spiral is only weakly formed,  when there are several
different spirals in the unit cell, or indeed when this picture
actually does not make the correct prediction.\cite{MKABH,RFMO,TAK}  
In early studies on Ni$_3$V$_2$O$_8$\cite{GL} and on
TbMnO$_3$\cite{MK} a phenomenological Landau theory was
developed which invoked a trilinear magnetoelectric interaction.
The virtue of this theory was that it showed exactly how the
crystal structure controlled the direction of the spontaneous
polarization, a phenomenon which had not been understood prior to
Refs. \onlinecite {GL}, \onlinecite{MK}, and \onlinecite{RFMO}.
Perhaps due to its simplicity, the subsequent Mostovoy\cite{MOSTOVOY} 
picture has been frequently applied.  However, as we shall
see here, in the present more complicated situation, this easily
visualized picture is hard to apply.  Accordingly, in the case we
discuss here concerning the switching properties\cite{HONDA}
in Mn$_2$GeO$_4$ (MGO)\cite{MGO1,MGO2,MGO3} the phenomenological
approach comes into its own.
Since the application of Landau theory\cite{ABH} is not trivial
in this case, the present paper will give a detailed explanation
as to how such a phenomenological theory is developed for MGO
to characterize the complex switching phenomena that occur at temperatures
below 5.5K.\cite{MGO1}

In Sec. II, we discuss how order parameters are introduced
to characterize the magnetic structure 
when it is nontrivial, having several magnetic 
sublattices of noncollinear
spins. The magnetic order at zero wave vector\cite{MGO1}
is characterized by order parameters of a standard type and
requires only a brief discussion.  We show how the
incommensurate magnetic ordering throughout the crystal
is specified in terms of a normalized
wave function which gives the distribution of magnetization
within a unit cell and the associated amplitudes which give rise
to complex-valued order parameters.  This description is based
on the mode structure and therefore provides a convenient
description of the symmetry of the
complicated incommensurate magnetic structure.
It is necessary, of course, to appropriately describe the
wave functions and order parameters of the different possible
domains consistent with the star of the wave vector.
The discussion of the incommensurate magnetic modes is
carried out mostly in an appendix and we will rely
on the symmetry analysis of the suite of computer programs
``ISODISTORT,''\cite{ISO} (whose results we verified by hand)
use of which automatically performs the symmetry analysis of the
magnetic ordering into sublattices.  We use the present example to
illustrate the use of this program, whose output is not
guaranteed to be in a form most conveniently suited to
an order parameter analysis.  

In Sec. III we discuss how the order parameters and the wave functions
for the various modes transform under the symmetry operations
of the crystal.\cite{HKAE}  In the case of MGO one has two possible wave-vector
domains, characterized by ${\bf k}_A = (k_x,k_y,0)$ and $-{\bf k}_A$,
and the other by ${\bf k}_B = (k_x,-k_y,0)$ and $-{\bf k}_B$.
Since the order parameters
incorporate the symmetry properties of the incommensurate magnetic
structure, they provide a natural way to discuss the properties of
this complex magnetic system.  Because the order parameters are
explicitly defined in terms of the magnetic modes, one can identify
transformations of the order parameters with actual transformation of
the magnetic structure.

Having analyzed how the order parameters transform under the
symmetry operations of the crystal, we construct, in Sec. IV, the
invariants which make up the magnetic free energy.
In principle a Landau theory could be developed to discuss the
ordering process.  However, this program is not our main objective
because the first-order transition at $T=5.5$K is quite complex.
Instead, we describe here the construction of the invariant
potentials which were used in Ref. \onlinecite{HONDA} to explain
the switching processes which take the system from one domain
structure to another.  The methodology used here may well be useful
in the description of other complex incommensurate magnetic systems.

\section{CRYSTAL SYMMETRY AND MODES}

\subsection{CRYSTAL SYMMETRY}

For these discussions we record the symmetry operations of
the orthorhombic space group for MGO, namely Pnma = No. 62 in 
Ref. \onlinecite{ITC},
where $x$, $y$, and $z$ refer to the $a$, $b$, and $c$ crystal axes:
\begin{eqnarray}
{\cal E} &=& (x,y,z) \ , \hspace{1.8 in} {\cal I} = (\overline x,
\overline y , \overline z ) \nonumber \\
m_x &=& (\overline x+ \oh ,y+\oh,z+\oh) \ , \hspace{0.4 in}
2_x = (x+\oh, \overline y +\oh , \overline z+\oh )
\nonumber \\
m_y &=& (x, \overline y +\oh ,z) \ , \hspace{1.1 in} 
2_y = (\overline x, y+\oh, \overline z) \nonumber \\
m_z &=& (x+\oh,y,\overline z+\oh) \ , \hspace{0.8 in} 
2_z = (\overline x+\oh , \overline y, z+\oh) \ .
\label{SYMOPS} \end{eqnarray}
We will also need to refer to the inverse operations:
\begin{eqnarray}
m_x^{-1} &=& (\overline x + \oh ,y-\oh,z-\oh) \ , \hspace{0.4 in}
2_x^{-1} = (x-\oh, \overline y +\oh , \overline z+\oh )
\nonumber \\
m_y^{-1} &=& (x, \overline y +\oh ,z) \ , \hspace{1.1 in} 
2_y^{-1} = (\overline x, y-\oh, \overline z) \nonumber \\
m_z^{-1} &=& (x-\oh,y,\overline z+\oh) \ , \hspace{0.8 in} 
2_z^{-1} = (\overline x+\oh , \overline y, z-\oh) \ .
\end{eqnarray}
The eight sites in the unit cell are:
\begin{eqnarray}
&& \tauv_1 = (0,0,0) \ , \hspace {1.4 in}
\tauv_2 = (1/2, 0 , 1/2 ) \nonumber \\
&& \tauv_3 = (1/2 , 1/2 , 1/2) \ , \hspace{1 in}
\tauv_4 = (0,1/2,0) \nonumber \\
&& \tauv_5 = (a,1/4,\epsilon)  \ , \hspace{1 in}
\tauv_6 = (a+1/2,1/4,1/2-\epsilon ) \nonumber \\ 
&& \tauv_7 = (1-a,3/4,-\epsilon) \ , \hspace{ 1 in}
\tauv_8 = (1/2-a,3/4,1/2+\epsilon)  \ ,
\label{TAUEQ} \end{eqnarray}
where $\epsilon\approx 0.0$ and $a \approx 0.275$.

\subsection{ZERO WAVE VECTOR MODES}

\begin{table}
\caption{\label{ZERO} Symmetry properties of the two active
zero-wave-vector magnetic order parameters $X_n$ for the irrep
$\Gamma_n$.  Both these order parameters (OP) are odd under time reversal.}
\vspace*{0.2 in}
\begin{tabular} {|c | c c c |}
\hline \hline
\ \ OP\ \ & $m_x$ & $m_y$ & $m_z$ \\
\hline
$X_1$ & $1$ & $1$ & $1$ \\
$X_3$ & \ \ $-1$\ \ &\ \ $-1$\ \ &\ \ $+1$\ \ \\
\hline \hline
\end{tabular}
\end{table}
Two zero wave vector irreducible representations (irreps)
are active: $\Gamma_1$ and $\Gamma_3$.
Their parity under the mirror operations is given in Table \ref{ZERO}.
The actual wave functions for these modes, given in the
Supplemental Material to Ref. \onlinecite{MGO1}, are
not needed for our symmetry analysis. However, it is helpful to note that
the phase with $X_1$ is a type of antiferromagnetic ordering and $X_3$ ordering
has a net magnetic moment along the $z$-axis.

\subsection{INCOMMENSURATE MODES}

In Appendix A, based on results from ISODISTORT\cite{ISO} we show that
the magnetization throughout a domain of wave vector ${\bf k}_A$
for the irrep $D^{(\sigma=1)}$ can be written as
\begin{eqnarray}
M_\alpha^{(A,1)}({\bf N}+\tauv_1) &=& a_\alpha 
e^{-i[{\bf N}+ \tauv_1] \cdot {\bf k}_A+i \phi } + {\rm c. \ c. } \nonumber  \\
M_\alpha^{(A,1)}({\bf N}+\tauv_2) &=& \mu_\alpha 
a_\alpha e^{-i[{\bf N}+ \tauv_2] \cdot {\bf k}+i \phi } + {\rm c. \ c. } 
\nonumber \\
M_\alpha^{(A,1)}({\bf N}+\tauv_3) &=&
b_\alpha e^{-i[{\bf N}+ \tauv_3] \cdot {\bf k}+i \phi } + {\rm c. \ c. } 
\nonumber \\
M_\alpha^{(A,1)}({\bf N}+\tauv_4) &=& \mu_\alpha
b_\alpha e^{-i[{\bf N}+ \tauv_4] \cdot
{\bf k}+i \phi } + {\rm c. \ c. } \nonumber \\
M_\alpha^{(A,1)}({\bf N}+\tauv_5) &=& 
z_\alpha e^{-i[{\bf N}+ \tauv_5] \cdot
{\bf k}+i \phi } + {\rm c. \ c. } \nonumber \\
M_\alpha^{(A,1)}({\bf N}+\tauv_6) &=& \mu_\alpha
z_\alpha e^{-i[{\bf N}+ \tauv_6] \cdot
{\bf k}+i \phi } + {\rm c. \ c. } \nonumber \\
M_\alpha^{(A,1)}({\bf N}+\tauv_7) &=& 
z_\alpha^* e^{-i[{\bf N}+ \tauv_7] \cdot
{\bf k}+i \phi } + {\rm c. \ c. } \nonumber \\
M_\alpha^{(A,1)}({\bf N}+\tauv_8) &=& \mu_\alpha
z_\alpha^* e^{-i[{\bf N}+ \tauv_8] \cdot
{\bf k}+i \phi } + {\rm c. \ c. } \ ,
\label{ALLEQ} \end{eqnarray}
where ${\bf N} \equiv (N_x,N_y,N_z)$ specifies the integer coordinates
(we always use rlu) of the unit cell, $a_\alpha$ and $b_\alpha$ are
real-valued, $z_\alpha$ is complex-valued,  $\mu_x=\mu_y=-\mu_z = -1$,
the superscripts label the wave vector and the irrep, and the
subscript is the component label, $\alpha =x, y,$ or $z$.
For economy in notation we write Eq. (\ref{ALLEQ}) as
\begin{eqnarray}
M_\alpha^{(A,1)}( {\bf N+\tauv_n}) &=& e^{i \phi} \left[
a_\alpha, \mu_\alpha a_\alpha, b_\alpha, \mu_\alpha b_\alpha;
z_\alpha , \mu_\alpha z_\alpha , z_\alpha^* , \mu_\alpha z_\alpha^* \right]_n
e^{-i[{\bf N + \tau_n)\cdot{\bf k}_A} }+ {\rm c. \ c.} \ .
\label{NOTE} \end{eqnarray}
In this notation the magnetization for irrep $D^{(\sigma=2)}$ is given by
\begin{eqnarray}
M_\alpha^{(A,2)} ( {\bf N+\tauv_n}) &=& e^{i \phi} \left[
c_\alpha, - \mu_\alpha c_\alpha, d_\alpha, - \mu_\alpha d_\alpha;
w_\alpha , - \mu_\alpha w_\alpha , w_\alpha^* , - \mu_\alpha w_\alpha^*
\right]_n e^{-i[{\bf N + \tau_n)\cdot{\bf k}_A} }+ {\rm c. \ c.} \ .
\label{MODEQ2} \nonumber \\
\end{eqnarray}
Note that we use different constants for irreps 1 and 2.  We do this
to emphasize the fact that the wave functions for different
symmetries are not related, just as atomic s and p functions are
not related to each other by symmetry.

To describe the magnetization distribution for a single irrep requires
specifying the 13 real-valued parameters, namely $a_\alpha$, $b_\alpha$,
$\Re (z_\alpha)$, $\Im (z_\alpha)$ for $\alpha=x,y,z$ and the global phase
$\phi$. A similar analysis was given in the Supplemental Material to Ref.
\onlinecite{MGO1}, but it appears to allow more adjustable parameters
than those given here, although our set of parameters is allowed by their
analysis. There the actual values of the parameters obtained from experiment
are given, but they are not needed here.

\subsection{DEFINITION OF ORDER PARAMETERS}

In the simplest situation only a single irrep is activated. In that case,
the above description of the incommensurate
magnetization holds throughout the range of existence of the
phase (i. e. as long as no phase boundary has been crossed), but the parameters
of the wave function (i. e. the $a$'s, $b$'s and $z$'s) depend on
temperature. If incommensurate magnetic ordering
appears below a continuous phase transition, then the 
temperature-dependence of the wavefunction when the ordering
initially develops gives rise mainly to a change of scale of
the coefficients.  Accordingly, we describe the wave function
as an amplitude times a normalized wave function.  In so doing,
we incorporate the phase factor $\exp (i \phi)$ in the
amplitude, thus giving rise to a complex-valued amplitude
known as the order parameter, here denoted ${\bf Q}^{(\sigma)}_X$,
for the irrep $\sigma$ at wave vector ${\bf k}_X$, where 
$X=\pm A$ or $X=\pm B$ (and ${\bf k}_{-X}$ denotes $-{\bf k}_X$).
Of course, it is an approximation to assume that the temperature
dependence of the wave function merely induces a temperature
dependence of the order parameter.  However, since this approximation
correctly describes the symmetry of the phase, it is often useful
and can form the basis for a renormalization group treatment of
the critical behavior.\cite{RG}
Furthermore, as shown in Appendix B, corrections due to the
additional temperature dependence of the wave function can be generated
within the Landau formulation of the order parameter which
we describe here. 

In the present situation we are interested in describing a system
which can have both irreps simultaneously present. This means that order
parameters for the two irreps can simultaneously be nonzero.
In what follows each mode is characterized by its order parameter
${\bf Q}^{(\sigma)}_{\bf X}$ which has its own magnitude and phase.
Thus we write the contribution to the $\alpha$-component of the
magnetization from irrep 1 at wave vector ${\bf k}_A$ to be
\begin{eqnarray}
M_\alpha^{(A,1)}( {\bf N} + \tauv_n) &=& {\bf Q}_A^{(1)} \left[
a_\alpha, \mu_\alpha a_\alpha, b_\alpha, \mu_\alpha b_\alpha;
z_\alpha , \mu_\alpha z_\alpha , z_\alpha^* , \mu_\alpha z_\alpha^* \right]_n
e^{-i {\bf k}_A \cdot [{\bf N} + \tauv_n]} + {\rm c. \ c.} \nonumber \\
& \equiv& {\bf Q}_A^{(1)} \Psi^{(A,1)}_{\alpha ,n} 
e^{-i {\bf k}_A \cdot [{\bf N} + \tauv_n]} + {\rm c. \ c.} 
\label{MODEEQ1} \end{eqnarray}
and that from irrep  2 at wave vector ${\bf k}_A$ to be
\begin{eqnarray}
M_\alpha^{(A,2)}( {\bf N} + \tauv_n) &=& {\bf Q}_A^{(2)} \left[
c_\alpha, -\mu_\alpha c_\alpha, d_\alpha, -\mu_\alpha d_\alpha;
w_\alpha , -\mu_\alpha w_\alpha , w_\alpha^* ,- \mu_\alpha w_\alpha^* \right]_n
e^{-i{\bf k}_A \cdot [{\bf N} + \tauv_n]} + {\rm c. \ c.} \nonumber \\ 
& \equiv& {\bf Q}_A^{(2)} \Psi_{\alpha,n}^{(A,2)}
e^{-i {\bf k}_A \cdot [{\bf N} + \tauv_n]} + {\rm c. \ c.} \ , 
\label{MODEEQ2} \end{eqnarray}
where we require the wave functions $\Psiv^{(A,m)}$ to be normalized:
\begin{eqnarray}
\sum_\alpha \left[ 2 a_\alpha^2 + 2 b_\alpha^2 + 4|z_\alpha|^2 \right] =
\sum_\alpha \left[ 2 c_\alpha^2 + 2 d_\alpha^2 + 4|w_\alpha|^2 \right] = 1 \ .
\end{eqnarray}
We can equally well write the equation for $M_\alpha^{(A,1)}({\bf N}+\tauv_n)$
as
\begin{eqnarray}
M_\alpha^{(A,1)}( {\bf N} + \tauv_n) &=& {{\bf Q}_A^{(1)}}^* \left[
a_\alpha, \mu_\alpha a_\alpha, b_\alpha, \mu_\alpha b_\alpha;
z_\alpha^* , \mu_\alpha z_\alpha^* , z_\alpha , \mu_\alpha z_\alpha \right]_n
e^{-i (-{\bf k}_A) \cdot [{\bf N} + \tauv_n]} + {\rm c. \ c.} \ .
\end{eqnarray}
This leads us to write
\begin{eqnarray}
{\bf Q}_{-X}^{(\sigma)} &=& {{\bf Q}_{X}^{(\sigma)}}^* \ , \ \ \
\Psi^{(-X,1)}_{\alpha ,n} = {\Psi^{(X,1)}_{\alpha ,n}}^* \ .
\label{XINDEX} \end{eqnarray}
where the script $-X$ refers to the wave vector $-{\bf k}_X$.  We assume
that the $\Psi_{\alpha n}^{(A,\sigma)}$ have been determined by fitting
experimental data, as in Ref. \onlinecite{MGO1}.  Our aim is to obtain the
trnasformation properties of the order parameteters and to determine the
wave functions for $\pm {\bf k}_B$ from those of $\pm {\bf k}_A$.

The results for $\Psiv$ for the star of the wave vector will be collected
in Table \ref{STAR}.  One can easily see that these two wave functions,
$\Psiv^{(\sigma=1)}$ and $\Psiv^{(\sigma=2)}$, are orthogonal to one
another.  We should point out that there is some arbitrariness in choosing
the sign of the order parameter ${\bf Q}$.  We could have chosen
$\Psiv$ to be the negative of that listed in Table \ref {STAR}.  This would
induce a change of sign in the associated ${\bf Q}$ and would give rise to
an equally valid representation.  This arbitrariness is also evident
in the case of a two-sublattice antiferromagnet, where we arbitarily choose
the orientations of the sublattice magnetizations when the order parameter
(the staggered magnetization) is positive.

\begin{table}
\caption{\label{STAR} Wave functions
$\Psi^{(X,\sigma)}_{\alpha n}$ for the star of the wave vector,
where $X=A$ or $X=B$ and
$\mu_\alpha=(-1,-1,+1)$ and $\lambda_\alpha'=(-1,+1,-1)$.
The notation is as in Eq. (\ref{MODEEQ1}).
Here $a_\alpha$, $b_\alpha$, $c_\alpha$, and $d_\alpha$ are real-valued and
$z_\alpha$ and $w_\alpha$ are complex-valued.  The values of these
parameters are not fixed by symmetry. The wave function for ${\bf k}_B$
is of the form given in Eq. (\ref{FORMOFPSI}). Its explicit relation to
that for ${\bf k}_A$ (given here) is obtained in Eqs.
(\ref{EQPSIB1}) and (\ref{EQPSIB2}).}

\vspace*{0.2 in}
\begin{tabular} {|c|cccccccc|}
\hline \hline
& $n=1$ & $n=2$ & $n=3$ & $n=4$ & $n=5$ & $n=6$ & $n=7$ & $n=8$ \\ 
\hline
\ \ $\Psi^{(A,1)}_{\alpha n}$ = \ \ &\ \ $a_\alpha$\ \ &\ \ 
$\mu_\alpha a_\alpha$\ \ &\ \ $b_\alpha$\ \ &\ \ $\mu_\alpha b_\alpha$\ \
& \ \ $z_\alpha$\ \ &\ \ $\mu_\alpha z_\alpha$\ \ 
&\ \ $z_\alpha^*$\ \ &\ \ $\mu_\alpha z_\alpha^*$\ \ \\
\ \ $\Psi^{(A,2)}_{\alpha n}$ = \ \ &\ \ $c_\alpha$\ \ &\ \ 
$- \mu_\alpha c_\alpha$\ \ &\ \ $d_\alpha$\ \ &\ \ $- \mu_\alpha d_\alpha$\ \
& \ \ $w_\alpha$\ \ &\ \ $- \mu_\alpha w_\alpha$\ \ 
&\ \ $w_\alpha^*$\ \ &\ \ $- \mu_\alpha w_\alpha^*$\ \ \\
\hline
\ \ $\Psi^{(B,1)}_{\alpha n}/\lambda_\alpha'$ = \ \ &
\ \ $\mu_\alpha b_\alpha$\ \ &\ \ $b_\alpha$\ \ &\ \ $\mu_\alpha a_\alpha$
\ \ &\ \ $a_\alpha$\ \ & \ \ $z_\alpha$\ \ &\ \ $\mu_\alpha z_\alpha$\ \ 
&\ \ $z_\alpha^*$\ \ &\ \ $\mu_\alpha z_\alpha^*$\ \ \\
\ \ $\Psi^{(B,2)}_{\alpha n}/\lambda_\alpha'$ = \ \ &
\ \ $\mu_\alpha d_\alpha$\ \ &\ \ $-d_\alpha$\ \ &\ \ $\mu_\alpha c_\alpha$
\ \ &\ \ $-c_\alpha$\ \ & \ \ $-w_\alpha$\ \ &\ \ $\mu_\alpha w_\alpha$\ \ 
&\ \ $-w_\alpha^*$\ \ &\ \ $\mu_\alpha w_\alpha^*$\ \ \\
\hline \hline
\end{tabular}
\end{table}

Usually the absolute phase of an order parameter is not important.
However, relative phases of order parameters can crucially affect
observable quantities, such as the electric polarization.  Also
note that we have made an arbitrary choice to associate ${\bf Q}_A^{(1)}$ 
with  $\exp (-i[N_x k_x + N_y k_y ])$ rather than
with $\exp (i[N_x k_x + N_y k_y ])$. 

\section{TRANSFORMATION PROPERTIES OF THE ORDER PARAMETERS}
\label{TRANSSUB}

We define the transformation of order parameters under an operator
${\cal O}$, by considering the effect of ${\cal O}$ on the distribution
of magnetization over the system.  We write
\begin{eqnarray}
{\bf M} ( N_x, N_y, N_z; \tauv_n )' &=& 
{\cal O}^S {\bf M} \left( [{\cal O}^R]^{-1}
[N_x, N_y, N_z; \tauv_n ] \right)  \ ,
\label{TRANSEQ} \end{eqnarray}
where ${\cal O}^S$ is the part of the operator ${\cal O}$ that operates on
spin and ${\cal O}^R$ is the part of the operator ${\cal O}$ that
operates on the position of the spin.  This equation says that the
transformed magnetic moment (indicated by a prime) that was at 
$\left[ {\cal O}^R \right]^{-1} [ {\bf N} + \tauv_n]$
has been transformed by  ${\cal O}^S$ and is placed at its final
location at ${\bf N} + \tauv_n$. In this section we will consider the
transformations under the three perpendicular mirror planes, since these
operations can be taken to be the generators of the point group.
In the course of this program we will identify the ``coordinaate system''
or ``unit vectors'' which in this case is the set of wave functions for each
wave vector. In Eqs. (\ref{MODEEQ1}) and (\ref{MODEEQ2}) we have already defined the
wave function for the wave vector ${\bf k}_A$ and in Eq. (\ref{XINDEX}) for the
wave vector $-{\bf k}_A$. In subsection IIIC we will obtain those for wave
vectors $\pm {\bf k}_B$.

\subsection{TRANSFORMATION BY INVERSION ${\cal I}$}

Perhaps the simplest operation is spatial inversion ${\cal I}$.
Since ${\cal I}^{-1}={\cal I}$, Eq. (\ref{TRANSEQ}) is
\begin{eqnarray}
{\bf M} ( N_x, N_y, N_z; \tauv_n )' &=& 
{\cal I}^S {\bf M} \left( [{\cal I}^R]
[N_x, N_y, N_z; \tauv_n ] \right)  
= {\bf M} \left( [{\cal I}^R]
[N_x, N_y, N_z; \tauv_n ] \right)  \ ,
\end{eqnarray}
where we used the fact that the magnetic moment is a pseudovector
to write the second version of the above equation. Thus, for
wave vector ${\bf k}_A$ and irrep $\sigma$
\begin{eqnarray}
{\bf M}_\alpha (N_x,N_y,N_z;\tauv_n)'&=& 
Q_A^{(\sigma)}{\cal I}^R [\Psi_{\alpha n}^{(A,\sigma)}
e^{-i{\bf k}_A \cdot ({\bf N} + \tauv_n)}] + {\rm c. c.} 
\nonumber \\
&=& Q_A^{(\sigma)}[{\cal I}^R \Psi_{\alpha n}^{(A,\sigma)}]
e^{-i[-{\bf k}_A \cdot ({\bf N}+\tauv_n)]} + {\rm c. \ c. }   \ .
\end{eqnarray}

\begin{table}
\caption{\label{ITAB} Transformation of $\tauv_n$ by $\cal I$. Since 
the vector $\tauv$ is not changed by adding a lattice vector to it, the
third column is equivalent to the fourth column.}
\vspace*{0.2 in}
\begin{tabular} {|ccccc|}
\hline \hline
$n$ & $\tauv_n$ & ${\cal I} \tauv_n$ & $\Rightarrow$ & $\tauv_n'$ \\
\hline
1 & $(0,0,0)$ &\ \ $(0, 0, 0)$\ \ &\ \
$(0,0,0)$\ \ &\ \ $\tauv_1$\ \  \\
2 & $(\frac{1}{2},0,\frac{1}{2})$ & $ (-\frac{1}{2},0,-\frac{1}{2})$ &
$(\frac{1}{2},0,\frac{1}{2}) $ & $\tauv_2$ \\
3 & $( \frac{1}{2}, \frac{1}{2}, \frac{1}{2})$ & $(-\frac{1}{2},-\frac{1}{2},-\frac{1}{2})$ &
$(\frac{1}{2},\frac{1}{2},\frac{1}{2})$ &  $\tauv_3$ \\
4 & $(0, \frac{1}{2},0)$ & $(0,-\frac{1}{2}, 0)$ &
$(0,\frac{1}{2},0)$ & $\tauv_4$ \\
5 & $(a, \frac{1}{4}, \epsilon )$ &\ \ $(-a, -\frac{1}{4}, -\epsilon)$\ \ &
\ \ $(1-a , \frac{3}{4},-\epsilon )$ & $\tauv_7$  \ \\
6 &\ \ $(a+ \frac{1}{2}, \frac{1}{4},\frac{1}{2}-\epsilon)$\ \ & 
$(-a-\frac{1}{2}, -\frac{1}{4},-\frac{1}{2}+\epsilon)$ 
& $(\frac{1}{2}-a, \frac{3}{4},\frac{1}{2}+\epsilon)$ & $\tauv_8$ \\
7 & $(1-a, \frac{3}{4}, -\epsilon )$ & $(a-1, -\frac{3}{4}, \epsilon)$  &
$(a, \frac{1}{4}, \epsilon)$ & $\tauv_5$ \\
\ 8\ & $( \frac{1}{2}-a, \frac{3}{4}, \frac{1}{2}+\epsilon )$ &
 $(a - \frac{1}{2},-\frac{3}{4},-\frac{1}{2}-\epsilon )$ & 
$ (a+\frac{1}{2}, \frac{1}{4},\frac{1}{2}-\epsilon )$ & $\tauv_6$ \\
\hline \hline
\end{tabular}
\end{table}

The effect of ${\cal I}$ on $\tauv_n$ is given in Table \ref{ITAB}:
\begin{eqnarray}
{\cal I}^R \Psi_{\alpha n}^{(A,\sigma)}= \Psi_{\alpha \overline n}^{(A,\sigma)}
= [ \Psi_{\alpha n}^{(A,\sigma)}]^* \ ,
\end{eqnarray}
where Eqs. (\ref{MODEEQ1}) and (\ref{MODEEQ2}) indicate that
$\Psi_{\alpha \overline n}^{(A,\sigma)} = [ \Psi_{\alpha n}^{(A,\sigma)}]^*$.  Thus,
\begin{eqnarray}
{\bf M}_\alpha (N_x,N_y,N_z;\tauv_n)' &=& {\bf Q}_A^{(\sigma)} [ \Psi_{\alpha n}^{(A,\sigma)}]^*
[ e^{-i [-{\bf k}_A \cdot ({\bf N}+\tauv_n)}] + {\rm c.\ c.} \ .
\end{eqnarray}
This is of the form
\begin{eqnarray}
{\bf M}_\alpha (N_x,N_y,N_z;\tauv_n)' &=& {{\bf Q}_{-A}^{(\sigma)}}'
\Psi_{\alpha n}^{(-A,\sigma)}
[ e^{-i [-{\bf k}_A \cdot ({\bf N}+\tauv_n)}] + {\rm c.\ c.} \ ,
\end{eqnarray}
with
\begin{eqnarray}
{{\bf Q}_{-A}^{(\sigma)}}' &\equiv& {\cal I} {\bf Q}_{-A}^{(\sigma)} = {{\bf Q}_A^{(\sigma)}} 
= \left[ {\bf Q}_{-A}^{(\sigma)} \right]^* \ ,
\ \ \ \ \ \ \ {\Psi_{\alpha n}^{(-A,\sigma)}}=[\Psi_{\alpha n}^{(A,\sigma)}]^* \ ,
\end{eqnarray}
consistent with Eq. (\ref{XINDEX}). The analogous result holds for ${\bf k}_B$.
So in all we have
\begin{eqnarray}
{\cal I} {\bf Q}^{(\sigma)}_X &=& {{\bf Q}^{(\sigma)}_X}^* 
= {{\bf Q}_{-X}^{(\sigma)}} \ .
\end{eqnarray}
We may consider the order
parameter to be a four-component column vector $\vec {\bf Q}_\sigma$ with components
$(Q_\sigma)_1 = Q_A^{(\sigma)}$, $(Q_\sigma)_2 = Q_B^{(\sigma)}$,
$(Q_\sigma)_3 = {Q_A^{(\sigma)}}^*$, $(Q_\sigma)_4 = {Q_B^{(\sigma)}}^*$.
We summarize our results for the effect of inversion by writing
\begin{eqnarray}
{\cal I} \vec{\bf Q}_\sigma &\equiv &
{\cal I} \left[ \begin{array} {c }  {\bf Q}_A^{(1)} \\ {\bf Q}_B^{(1)} \\
{\bf Q}_{-A}^{(1)} \\
{\bf Q}_{-B}^{(1)}  \\ \end{array} \right] = \left[ \begin{array} {c c c c}
0 & 0 & 1 & 0 \\ 0 & 0 & 0 & 1 \\ 
1 & 0 & 0 & 0 \\ 0 & 1 & 0 & 0 \\ 
\end{array} \right]
\left[ \begin{array} {c } {\bf Q}_A^{(1)} \\ {\bf Q}_B^{(1)} 
\\ {\bf Q}_{-A}^{(1)} \\ {\bf Q}_{-B}^{(1)} \\ \end{array} \right] 
\equiv {\cal M}_\sigma({\cal I}) {\vec {\bf Q}}_\sigma \ .
\label{EQ20} \end{eqnarray}
In general, if ${\cal O}$ is an operator, then we write
\begin{eqnarray}
{\cal O} (Q_\sigma)_n &=& \sum_m [ {\cal M}_\sigma ( {\cal O})]_{nm} (Q_\sigma )_m \ .
\end{eqnarray}

\subsection{TRANSFORMATION BY $m_z$}

Since $m_z$ leaves the wave vector ${\bf k}$ invariant, we consider it next.
We start by considering the case when ${\bf k}={\bf k}_A$.  Thus we apply
Eq. (\ref{TRANSEQ}) when ${\bf k}={\bf k}_A$ and ${\cal O}=m_z$:
\begin{eqnarray}
M_\alpha ( N_x, N_y, N_z; \tauv_n )' &=& 
\lambda_\alpha M_\alpha \left( [{m_z}^R]^{-1} [N_x, N_y, N_z; \tauv_n ]
\right)  \ ,
\end{eqnarray}
where, since ${\bf M}$ is a pseudovector, $\lambda_\alpha=(-1,-1,+1)$.
Thereby we find that
\begin{eqnarray}
M_\alpha ( N_x, N_y, N_z; \tauv_n )' &=& 
\lambda_\alpha \left( [m_z^R]^{-1} \Psi_{\alpha n}^{(A,1)} \right)
e^{-i {\bf k}_A \cdot [m_z^R]^{-1} [ {\bf N} + \tauv_n]} {\bf Q}_A^{(1)} 
 + {\rm c. \ c.} \ .
\end{eqnarray}
To evaluate the exponential for $k_z=0$, note that acting on a vector of
the form $(v_x,v_y,0)$, we have
$[m_z^R]^{-1}(v_x,v_y,0)=(v_x-1/2,v_y,0)$. So
\begin{eqnarray}
M_\alpha ( N_x, N_y, N_z; \tauv_n )' &=& 
\lambda_\alpha \left( [m_z^R]^{-1} \Psi_{\alpha n}^{(A,1)}\right)
e^{-i {\bf k}_A \cdot [ {\bf N} + \tauv_n]} e^{ik_x/2} {\bf Q}_A^{(1)} 
+ {\rm c. \ c.} \ .
\end{eqnarray}
Now we consider $[m_z^R]^{-1} \Psi_{\alpha n}^{(A,1)}$.
In Table \ref{TAUTAB} we see that 
\begin{eqnarray}
[m_z^R]^{-1} \Psi_{\alpha n}^{(A,1)} &=&
\Psi_{\alpha \overline n}^{(A,1)} \ ,
\end{eqnarray}
where $\overline n = n-1$ if $n$ is even and $\overline n = n+1$ if $n$ is odd.
But, since $\mu_\alpha = 1/\mu_\alpha$,  we have, from Eq. (5), that
\begin{eqnarray}
\Psi_{\alpha \overline n}^{(A,1)}  = \mu_\alpha \Psi_{\alpha n}^{(A,1)} \ .
\label{EQ15} \end{eqnarray}
Note that $\mu_\alpha \lambda_\alpha=1$, so that the final result is
\begin{eqnarray}
M_\alpha ( N_x, N_y, N_z; \tauv_n )' &=& 
e^{ik_x/2} \Psi_{\alpha n}^{(A,1)}
e^{-i {\bf k}_A \cdot [ {\bf N} + \tauv_n]} {\bf Q}_A^{(1)} 
+ {\rm c. \ c.} \ .
\end{eqnarray}
This is of the form
\begin{eqnarray}
M_\alpha ( N_x, N_y, N_z; \tauv_n )' &=& 
{{\bf Q}_A^{(1)}}' {\Psi_{\alpha n}^{(A,1)}}
e^{-i {\bf k}_A \cdot [ {\bf N} + \tauv_n]} + {\rm c. \ c.} \ .
\end{eqnarray}
In other words,
\begin{eqnarray}
{{\bf Q}_A^{(1)}}' &=& m_z {\bf Q}_A^{(1)}= e^{ik_x/2} {\bf Q}_A^{(1)} \ .
%
\label{EQ24} \end{eqnarray}
For $\sigma=2$ we have
\begin{eqnarray}
m_z {\bf Q}_A^{(2)} &=& -e^{ik_x/2} {\bf Q}_A^{(2)} \ .
\end{eqnarray}
The difference in sign for 
$m_z {\bf Q}_A^{(2)}$ occurs because, here, instead of Eq. (\ref{EQ15}), one has
\begin{eqnarray}
\Psi_{\alpha \overline n}^{(A,2)}  = - \mu_\alpha \Psi_{\alpha n}^{(A,2)} \ .
\end{eqnarray}

\begin{table}
\caption{\label{TAUTAB} As table \ref{ITAB}. Transformation of $\tauv_n$ by $m_z$.}
\vspace*{0.2 in}
\begin{tabular} {|ccccc|}
\hline \hline
$n$ & $\tauv_n$ & $[m_z^R]^{-1} \tauv_n$ & $\Rightarrow$ & $\tauv_n'$ \\
\hline
1 & $(0,0,0)$ &\ \ $(- \frac{1}{2}, 0, \frac{1}{2})$\ \ &\ \
$(\frac{1}{2},0, \frac{1}{2})$\ \ &\ \ $\tauv_2$\ \  \\
2 & $(\frac{1}{2},0,\frac{1}{2})$ & $(0,0,0)$ & $(0,0,0)$ & $\tauv_1$ \\
3 & $( \frac{1}{2}, \frac{1}{2}, \frac{1}{2})$ & $(0,\frac{1}{2},0)$ &
$(0,\frac{1}{2},0)$ &  $\tauv_4$ \\
4 & $(0, \frac{1}{2},0)$ & $(-\frac{1}{2},\frac{1}{2}, \frac{1}{2})$ &
$(\frac{1}{2},\frac{1}{2},\frac{1}{2})$ & $\tauv_3$ \\
5 & $(a, \frac{1}{4}, 0 )$ &\ \ $(a-\frac{1}{2}, \frac{1}{4},\frac{1}{2})$\ \ &
\ \ $(a + \frac{1}{2}, \frac{1}{4},\frac{1}{2} )$ & $\tauv_6$  \ \\
6 &\ \ $(a+ \frac{1}{2}, \frac{1}{4},\frac{1}{2})$\ \ & $(a, \frac{1}{4},0)$ &
$(a, \frac{1}{4},0)$ & $\tauv_5$ \\
7 & $(1-a, \frac{3}{4}, 0 )$ & $(\frac{1}{2}-a$, $\frac{3}{4}$, $\frac{1}{2})$ &
$(\frac{1}{2}-a$, $\frac{3}{4}$, $\frac{1}{2})$ & $\tauv_8$ \\
\ 8\ & $( \frac{1}{2}-a$, $\frac{3}{4}$, $\frac{1}{2})$ & $(-a, \frac{3}{4},0)$ &
$(1-a, \frac{3}{4},0)$ & $\tauv_7$ \\
\hline \hline
\end{tabular}
\end{table}

We now use the above results to obtain analogous results for
wave vectors ${\bf k}_B$. Since the value of ${\bf k}$
does not appear explicitly, the wave functions of
${\bf k}_B=(k_x,-k_y,0)$ are of the form
\begin{eqnarray}
\Psiv^{(B,1)}&=& [ a_\alpha' , \mu_\alpha a_\alpha' , 
b_\alpha' , \mu_\alpha b_\alpha',
z_\alpha' , \mu_\alpha z_\alpha' , z_\alpha^*, \mu_\alpha {z_\alpha'}^* ] \ 
\label{FORMOFPSI} \end{eqnarray}
and similarly for $\Psiv^{(B,2)}$.
The relation between ($a'$, $b'$, $z'$) and ($a$, $b$, $z$), given in
Table \ref{STAR} will be derived later. Note that Eq. (\ref{EQ24}) holds 
when ${\bf k}_A$ is replaced by ${\bf k}_B$. Thus we obtain
\begin{eqnarray}
m_z {\vec {\bf Q}}_\sigma  = m_z \left[ \begin{array} {c }  
{\bf Q}_A^{(\sigma)} \\ {\bf Q}_B^{(\sigma)} \\ 
{\bf Q}_{-A}^{(\sigma)} \\ {\bf Q}_{-B}^{(\sigma)}  \\ \end{array} \right]
= (-)^{\sigma+1} \left[ \begin{array} {c c c c}
e^{ik_x/2} & 0 & 0 & 0 \\ 0 & e^{ik_x/2} & 0 & 0 \\ 
0 & 0 & e^{-ik_x/2} & 0 \\ 0 & 0 & 0 & e^{-ik_x/2} \\ \end{array} \right]
\left[ \begin{array} {c } {\bf Q}_A^{(\sigma)} \\ {\bf Q}_B^{(\sigma)} \\
{\bf Q}_{-A}^{(\sigma)} \\ {\bf Q}_{-B}^{(\sigma)} \\ \end{array} \right]
= {\cal M}_\sigma (m_z) {\vec {\bf Q}}_\sigma \ .
\label{TRANSZ} \end{eqnarray}
Here we noted that ${\bf Q}_{-X}^{(\sigma)} =[ {\bf Q}_{X}^{(\sigma)}]^*$
to obtain the lower half of the matrix.

\subsection{TRANSFORMATION BY $m_y$}
\label{IIIB}

To identify the modes for wave vector ${\bf k}_B$ from those of wave vector
${\bf k}_A$, we transform the wave functions for ${\bf k}_A$ into those
for ${\bf k}_B=(k_x,-k_y,0)$.  Although symmetry allows the parameters
of the wave function (e. g. $a$, $b$, etc.) to be arbitrary, once they are
fixed for wave vector ${\bf k}_A$, they are implicitly fixed (to within a
phase factor) for wave vector ${\bf k}_B$. Under
transformation by $m_y$ we write Eq. (\ref{TRANSEQ}) as
\begin{eqnarray}
M_\alpha ( N_x, N_y, N_z; \tauv_n )' &=& 
\lambda_\alpha' M_\alpha \left( [{m_y}^R]^{-1} 
[N_x, N_y, N_z; \tauv_n ] \right)  \ ,
\end{eqnarray}
where, since ${\bf M}$ is a pseudovector, $\lambda_\alpha'=(-1,+1,-1)$.
Thereby, for irrep $\sigma$  we find that
\begin{eqnarray}
M_\alpha ( N_x, N_y, N_z; \tauv_n )' &=& \lambda_\alpha' \left( [m_y^R]^{-1}
\Psi_{\alpha n}^{(A,\sigma)}\right) e^{-i {\bf k}_A \cdot [m_y^R]^{-1}
[ {\bf N} + \tauv_n]} {\bf Q}_A^{(\sigma)} + {\rm c. \ c.} \ ,
\end{eqnarray}
To evaluate the exponential for $k_z=0$, note that acting on a vector of
the form $(v_x,v_y,0)$, we have $[m_y^R]^{-1}(v_x,v_y,0)=(v_x,1/2-v_y,0)$.
So
\begin{eqnarray}
M_\alpha ( N_x, N_y, N_z; \tauv_n )' &=& \lambda_\alpha' 
\left( [m_y^R]^{-1} \Psi_{\alpha n}^{(A,\sigma)}\right)
e^{-i \left[ k_x ( N_x + \tau_{nx}) + k_y( -N_y -\tau_{ny} + 1/2) \right]}
{\bf Q}_A^{(\sigma)} + {\rm c. \ c.} \ .
\end{eqnarray}
Now we consider $[m_y^R]^{-1} \Psi^{(A,\sigma)}_{\alpha n}$.  In Table 
\ref{TAUTABY} we see that
\begin{eqnarray}
[m_y^R]^{-1} \Psi_{\alpha n}^{(A,\sigma)} &=&
\Psi_{\alpha \overline n}^{(A,\sigma)} \ ,
\end{eqnarray}
where $\overline n = 5-n$ if $n<5$ and $\overline n = n$ for $n>4$.  Then
\begin{eqnarray}
M_\alpha ( N_x, N_y, N_z; \tauv_n )' &=& \lambda_\alpha' e^{-ik_y/2} 
\Psi_{\alpha \overline n}^{(A,\sigma)}
e^{-i \left[ k_x ( N_x + \tau_{nx}) + (-k_y)(N_y +\tau_{ny}) \right]}
{\bf Q}_A^{(\sigma)} + {\rm c. \ c.} \ .
\end{eqnarray}
This is of the form

\newpage
\begin{table} [h!]
\caption{\label{TAUTABY} As Table \ref{ITAB}. Transformation of $\tauv_n$ by $m_y$.}
\vspace*{0.2 in}
\begin{tabular} {|ccccc|}
\hline \hline
$n$ & $\tauv_n$ & $[m_y^R]^{-1} \tauv_n$ & $\Rightarrow$ & $\tauv_n'$ \\
\hline
1 & $(0,0,0)$ &\ \ $(0, \frac{1}{2}, 0)$\ \ &\ \
$(0,\frac{1}{2},0)$\ \ &\ \ $\tauv_4$\ \  \\
2 & $(\frac{1}{2},0,\frac{1}{2})$ & $(\frac{1}{2},\frac{1}{2},\frac{1}{2})$
& $(\frac{1}{2},\frac{1}{2},\frac{1}{2})$ & $\tauv_3$ \\
3 & $( \frac{1}{2}, \frac{1}{2}, \frac{1}{2})$ & $(\frac{1}{2},0,\frac{1}{2})$
& $(\frac{1}{2},0,\frac{1}{2})$ &  $\tauv_2$ \\
4 & $(0, \frac{1}{2},0)$ & $(0,0, 0)$ &
$(0,0,0)$ & $\tauv_1$ \\
5 & $(a, \frac{1}{4}, 0 )$ &\ \ $(a, \frac{1}{4},0)$\ \ &
\ \ $(a , \frac{1}{4}, 0)$ & $\tauv_5$  \ \\
6 &\ \ $(a+ \frac{1}{2}, \frac{1}{4},\frac{1}{2})$\ \ & 
$(a+\frac{1}{2}, \frac{1}{4},\frac{1}{2})$ &
$(a + \frac{1}{2},\frac{1}{4},\frac{1}{2})$ & $\tauv_6$ \\
7 & $(1-a, \frac{3}{4}, 0 )$ & $(1-a$, $-\frac{1}{4}$, $0)$ &
$(1-a$, $\frac{3}{4}$, $0$ & $\tauv_7$ \\
\ 8\ &\ \ $( \frac{1}{2}-a$, $\frac{3}{4}$, $\frac{1}{2}-a)$\ \ &\ \ 
$(\frac{1}{2}-a, -\frac{1}{4},\frac{1}{2}-a)$ \ \ &\ \
$(\frac{1}{2}-a, \frac{3}{4},\frac{1}{2}-a)$ & $\tauv_8$ \\
\hline \hline
\end{tabular}
\end{table}

\begin{eqnarray}
M_\alpha ( N_x, N_y, N_z; \tauv_n )' &=& {{\bf Q}_B^{(\sigma)}}' \Psi_{\alpha n}^{(B,\sigma)}
e^{-i \left[ k_x ( N_x + \tau_{nx}) + (-k_y)(N_y +\tau_{ny}) \right]} + {\rm c. \ c.} \ .
\end{eqnarray}
We choose the signs of the wave functions for ${\bf k}_B$ such that
\begin{eqnarray}
\Psi^{(B,1)}_{\alpha n} &=& \lambda_\alpha' \Psi_{\alpha \overline n}^{(A,1)}
= \lambda'_\alpha [ \mu_\alpha b_\alpha,
b_\alpha , \mu_\alpha a_\alpha , a_\alpha ;
z_\alpha , \mu_\alpha z_\alpha , z_\alpha^* , \mu_\alpha z_\alpha^* ]_n \ .
\label{EQPSIB1}\\
\Psi^{(B,2)}_{\alpha n} &=& - \lambda_\alpha' \Psi_{\alpha \overline n}^{(A,2)}
= \lambda'_\alpha [ \mu_\alpha d_\alpha,
-d_\alpha , \mu_\alpha c_\alpha , -c_\alpha ;
-w_\alpha , \mu_\alpha w_\alpha ,-w_\alpha^* , \mu_\alpha w_\alpha^* ]_n \ .
\label{EQPSIB2}\end{eqnarray}
As expected, $\Psi_{\alpha n}^{(B,1)}$ is of the form of Eq. (\ref{FORMOFPSI}),
but now we have an explicit evaluation of $\Psi_{\alpha n}^{(B,\sigma)}$,
given in Table \ref{STAR}.  With these definitions the transformed
value of the order parameter ${\bf Q}_B^{(\sigma)}$ is
\begin{eqnarray}
{{\bf Q}_B^{(\sigma)}}' = m_y {\bf Q}_B^{(\sigma)} = (-1)^{\sigma+1}
{\bf Q}_A^{(\sigma)} e^{-ik_y/2} \ .
\label{EQ42} \end{eqnarray}

We repeat our previous warning about the phase. We could have defined
$\Psi^{(B,\sigma )}_{\alpha n}$ to be the negative of its value in Eqs. 
(\ref{EQPSIB1}) or (\ref{EQPSIB2}). This possibility is analyzed in
Appendix C, where we see that the choice of sign for the wave function
implies a choice of sign for the order parameters, but does not affect the
invariant potentials determined below.

To summarize: in terms of the order parameter vector $\vec {\bf Q}_\sigma$
introduced in Eq. (\ref{EQ20}), we have  $(m_y \vec{\bf Q}_\sigma)_n=
\sum_m {\cal M}_\sigma (m_y)_{nm} (\vec{\bf Q}_\sigma )_m$, with
\begin{eqnarray}
{\cal M}_\sigma (m_y) &=& (-1)^{\sigma+1} \left[ \begin{array} {c c c c}
0 & e^{ik_y/2} & 0 & 0 \\ e^{-ik_y/2}  & 0 & 0 & 0 \\
0 & 0 & 0 & e^{-ik_y/2} \\
0 & 0 & e^{ik_y/2} & 0 \\
\end{array} \right] \ . 
\label{EQYMAT} \end{eqnarray}
Equation  (\ref{EQ42}) gives the 2,1 element of ${\cal M}_\sigma (m_y)$.  The other
matrix elements can be deduced by changing the sign of $k_y$ or by complex conjugation.

\subsection{TRANSFORMATION BY $m_x$}

We now consider transformation by $m_x$.
We write Eq. (\ref{TRANSEQ}) for irrep $\sigma$ as
\begin{eqnarray}
M_\alpha ( {\bf N} + \tauv_n)' &=&
m_x^S M_\alpha^{(A,\sigma)} \left( [m_x^R]^{-1} [N_x,N_y,N_z; \tauv_n]
\right) \ .
\label{EQA}
\end{eqnarray}
Since ${\bf M}$ is a pseudovector we set $m_x^S=\lambda_\alpha''$,
with $\lambda'' = (1,-1,-1)$. Then
\begin{eqnarray}
\label{EQC}
M_\alpha ( {\bf N} + \tauv_n)' &=&
\lambda_\alpha''  \left[ [m_x^R]^{-1} \Psi^{(A,\sigma)}_{\alpha n} \right]
e^{-ik_A\cdot [m_x^R]^{-1} [ {\bf N} + \tauv_n]} {\bf Q}_A^{(\sigma)} 
+ {\rm c. \ c.} \ .
\end{eqnarray}
To evaluate the exponential for $k_z=0$, note that acting on a vector
of the form $(k_x,k_y,0)$, we have $[m_x^R]^{-1} (v_x,v_y,0)=
(1/2-v_x, v_y-1/2,-1/2)$.  So
\begin{eqnarray}
M_\alpha (N_x,N_y,N_z;\tauv_n)' &=& \lambda_\alpha''
{\bf Q}_A^{(\sigma)} \Psi^{(A,\sigma)}_{\alpha, \overline n}
e^{-i[k_x(-N_x)+k_yN_y] - i[k_x (-\tau_{n,x}+1/2) +k_y(\tau_{n,y}-1/2)]}
+ {\rm c. \ c.} \nonumber \\
&=& \lambda_\alpha'' {\bf Q}_A^{(\sigma)} 
\Psi^{(A,\sigma)}_{\alpha,\overline n}
e^{-i[(-k_x) N_x+k_yN_y + (k_y\tau_{ny} - k_x \tau_{nx})
+ (k_x - k_y)/2]} \  + {\rm c. \ c.} \ ,
\label{EQD} \end{eqnarray}
where now $\tauv_{\overline n}=m_x^R \tauv_n$, so that 
\begin{eqnarray}
\overline 1&=&3 \ , \ \ \ \overline 2=4 \ , \ \ \
\overline 3=1 \ , \ \ \ \overline 4=2 \ , \ \ \
\overline 5=8 \ , \ \ \ \overline 6=7 \ , \ \ \
\overline 7=6 \ , \ \ \ \overline 8=5 \ .
\label{EQ47} \end{eqnarray}
Equation (\ref{EQD}) is of the form
\begin{eqnarray}
M_\alpha (N_x,N_y,N_z;\tauv_n)' &=& {{\bf Q}_{-B}^{(\sigma)}}'
e^{-i[-{\bf k}_B \cdot ({\bf N}+ \tauv_n)} \Psi^{(-B,\sigma)}_{\alpha n}
+ {\rm c. \ c.} .
\end{eqnarray}
Thus we have
\begin{eqnarray}
{{\bf Q}_{-B}^{(\sigma)}}' = m_x {\bf Q}_{-B}^{(\sigma)} &=& \rho_\sigma
{\bf Q}_A^{(\sigma)} e^{-i(k_x-k_y)/2} \ , 
\hspace{0.5 in}
\Psi^{(-B,\sigma)}_{\alpha n} = \rho_\sigma
\Psi^{(A,\sigma)}_{\alpha \overline n} \lambda''_\alpha \ .
\label{EQ49} \end{eqnarray}
The reason we have included the factor $\rho_\sigma = \pm 1$ is because
$\Psi^{(B,\sigma)}$ was already fixed by Eqs. (\ref{EQPSIB1})
and (\ref{EQPSIB2}).  Accordingly,
here we have to choose the sign of $\rho_\sigma$ to be consistent
with our previous definition  of $\Psi^{(B,\sigma)}$. Using Eq.
(\ref{EQ47}) and taking $\Psi_{\alpha n}^{(A,\sigma)}$ from Table 
\ref{STAR} we find that Eq. (\ref{EQ49}) gives
\begin{eqnarray}
\rho_1 \lambda_\alpha'' [ \Psi_{\alpha \overline n}^{(A,1)}]^* &=&
\Psi_{\alpha n}^{(B,1)} = {\Psi_{\alpha n}^{(-B,1)}}^*
\nonumber \\ &=&
\rho_1 \lambda_\alpha'' [b_\alpha, \mu_\alpha b_\alpha , a_\alpha,
\mu_\alpha a_\alpha ; \mu_\alpha z_\alpha , z_\alpha , 
\mu_\alpha z_\alpha^*, z_\alpha^* ] \label{EQ50} \\
\rho_2 \lambda_\alpha'' [ \Psi_{\alpha \overline n}^{(A,2)}]^* &=&
\Psi_{\alpha n}^{(B,2)} = {\Psi_{\alpha n}^{(-B,2)}}^*
\nonumber \\ &=&
\rho_2 \lambda_\alpha'' [d_\alpha, -\mu_\alpha d_\alpha , c_\alpha,
- \mu_\alpha c_\alpha ; -\mu_\alpha w_\alpha , w_\alpha , 
-\mu_\alpha w_\alpha^*, w_\alpha^* ] .  \label{EQ51}
\end{eqnarray}
Comparing Eqs. (\ref{EQPSIB1}) and (\ref{EQ50}) we require that
$\rho_1 \lambda_\alpha'' = \lambda_\alpha' \mu_\alpha$ which gives
$\rho_1=+1$. Comparing Eqs. (\ref{EQPSIB2}) and (\ref{EQ51}) we require that
$\rho_2 \lambda_\alpha'' = \lambda_\alpha' \mu_\alpha$ which gives
$\rho_2=+1$.  The final result is that in terms of the order parameter
vector ${\vec {\bf Q}}_\sigma$ introduced in Eq. (\ref{EQ20}), we have
$m_x {\vec{\bf Q}}_\sigma = {\cal M}_\sigma (m_x) {\vec {\bf Q}}_\sigma$, with
\begin{eqnarray}
{\cal M}_\sigma (m_x) &=&  \left[ \begin{array} {c c c c}
0 & 0 & 0 & e^{i(k_x+k_y)/2} \\ 0 & 0 & e^{i(k_x-k_y/2)}  & 0 \\
0 & e^{-i(k_x+k_y)/2} & 0 & 0 \\
e^{-i(k_x-k_y)/2} & 0 & 0 & 0 \\
\end{array} \right] \ . 
\label{TRANSX} \end{eqnarray}
In Eq. (\ref{EQ49}) we have explicitly calculated the (4,1) matrix element
of the matrix ${\cal M}_\sigma (m_x)$. The other matrix elements
can be obtained by suitably changing the sign(s) of the components of
the wave vector(s).

\subsection{TRANSFORMATION BY $2_z$}

We write for irrep $\sigma=1$ and wave vector ${\bf k}_A$ under transformation by
$2_z$
\begin{eqnarray}
M_\alpha (N_x,N_y,N_z; \tauv_n)' &=& \lambda_\alpha''' M_\alpha
\left( [2_z^R]^{-1} [ N_x,N_y,N_z; \tauv_n ] \right) \nonumber \\
&=& \lambda_\alpha''' \left( [2_z^R]^{-1} \Psi^{(A,1)}_{\alpha n} \right)
e^{-i{\bf k}_A \cdot [2_z^R]^{-1} [ {\bf N} + \tauv_n]}
{\bf Q}_A^{(1)} + {\rm c. \ c.} \nonumber \\
&=& \lambda_\alpha''' \Psi^{(A,1)}_{\alpha \overline n} 
e^{-ik_x[-N_x - \tau_{nx} + 1/2] -ik_y[ -N_y - \tau_{ny}]} 
{\bf Q}_A^{(1)} + {\rm c. \ c.} \nonumber \\
&=& \lambda_\alpha''' \Psi^{(A,1)}_{\alpha \overline n} 
e^{-i(-k_x)[N_x + \tau_{nx} - 1/2] -i(-k_y)[ N_y + \tau_{ny}]} 
{\bf Q}_A^{(1)} + {\rm c. \ c.} \ ,
\label{EQ53} \end{eqnarray}
where $\overline 1 =2$, $\overline 3 = 4$,
$\overline 5=8$, $\overline 6 =7$, and the inverse relations also hold, so
that $\Psi_{\alpha \overline n}^{(A,1)}= \mu_\alpha {\Psi_{\alpha n}^{(A,1)}}^*$.
Also $\lambda_\alpha'''=(-1,-1,+1)$, so that $\lambda_\alpha'''=\mu_\alpha$.
Thus Eq. (\ref{EQ53}) is of the form
\begin{eqnarray}
M_\alpha (N_x,N_y,N_z;\tauv_n)' &=& {{\bf Q}_{-A}^{(1)}}'
\Psi_{\alpha n}^{(-A,1)} e^{-i[(-k_x)(N_x+\tau_{nx})
+ (-k_y)(N_y+ \tau_{ny})]} \ ,
\end{eqnarray}
with
\begin{eqnarray}
{{\bf Q}_{-A}^{(1)}}' = 2_z {\bf Q}_{-A}^{(1)} = {\bf Q}_A^{(1)} e^{-ik_x/2} \ , \ \ \ \
\Psi^{(-A,1)}_{\alpha n} = \lambda_\alpha'''
\Psi^{(A,1)}_{\alpha \overline n} = {\Psi^{(A,1)}_{\alpha n}}^* \ ,
\end{eqnarray}
in agreement with Eq. (\ref{XINDEX}). Since ${\bf Q}_{-B}^{(1)}$ has the same value of $k_x$,
the above result implies that 
\begin{eqnarray}
{{\bf Q}_{-B}^{(1)}}' = 2_z {\bf Q}_{-B}^{(1)} = {\bf Q}_B^{(1)} e^{-ik_x/2} \ .
\end{eqnarray}
The transformation properties of ${\bf Q}_X^{(1)}$ are obtained from those of 
${\bf Q}_{-X}^{(1)}$ by changing the sign of ${\bf k}$.
A similar analysis for irrep 2 (but with $\Psi_{\alpha \overline n}^{(A,2)}=
- \mu_\alpha {\Psi_{\alpha n}^{(A,2)}}^*$) leads to the final result that
$2_z {\vec {\bf Q}}_\sigma = {\cal M}(2_z) {\vec {\bf Q}}_\sigma$,
with
\begin{eqnarray}
{\cal M}_\sigma (2_z) &=&
(-1)^{\sigma+1} \left[ \begin{array} {c c c c}
0 & 0 & e^{ik_x/2} & 0 \\ 0 & 0 & 0  & e^{ik_x/2}  \\
e^{-ik_x/2}  & 0 & 0 & 0 \\
0 & e^{-ik_x/2} & 0 & 0 \\ \end{array} \right] \ . 
\label{LHSEQ} \end{eqnarray}

\subsection{TRANSFORMATION BY OTHER OPERATIONS}

Here we record the result for translation.
For instance, apply Eq. (\ref{TRANSEQ}) to the magnetization
when the transformation operator
is a translation ${\bf T}$ through a lattice vector:
\begin{eqnarray}
[{\bf M}( {\bf N}+\tauv_n)]' &=& {\bf T}^S {\bf M} \left( [{\bf T}^R ]^{-1}
[{\bf N} + \tauv_n ] \right) \ ,
\end{eqnarray}
This gives, for translation ${\bf T}$:
\begin{eqnarray}
[{\bf M}({\bf N}+\tauv_n)]' &=& [{\bf M}({\bf N}-{\bf T}+\tauv_n)] \nonumber \\
&=& e^{i {\bf k} \cdot {\bf T}} [{\bf M}({\bf N} + \tauv_n)] \ .
\label{TEQ} \end{eqnarray}
When the translation ${\bf T}$ is through an integer number of lattice
constants in the three lattice directions, $N_x,N_y,N_z$, we write
\begin{eqnarray}
{\cal M}_\sigma ({\bf T}_{N_x,N_y,N_z}) &=&
\left[ \begin{array} {c c c c}
e^{i(k_xN_x+k_yN_y)} & 0 & 0 & 0 \\ 0 & e^{i(k_xN_x-k_yN_y)}  & 0  & 0  \\
0 & 0 & e^{i(-k_xN_x-k_yN_y)} & 0 \\
0 & 0 & 0 & e^{i(-k_xN_x +k_yN_y )} \\ \end{array} \right] \ . 
\end{eqnarray}

\subsection{COMPOSITION RULES}

If ${\cal O}^{(1)}$ and ${\cal O}^{(2)}$ are two operators, then we might write that
\begin{eqnarray}
{\cal O}^{(1)} {\cal O}^{(2)} \left( \vec{\bf Q}_\sigma \right)_n &\equiv&
{\cal O}^{(1)} \left[ {\cal O}^{(2)} \left( \vec{\bf Q}_\sigma \right)_n \right]
= {\cal O}^{(1)} \left[ \sum_m \left( {\cal M}_\sigma ({\cal O}^{(2)}) \right)_{nm}
\left( \vec{\bf Q}_\sigma \right)_m \right] \nonumber \\
&=& \sum_m \sum_s \left( {\cal M}_\sigma ({\cal O}^{(2)}) \right)_{nm} 
\left( {\cal M}_\sigma ({\cal O}^{(1)}) \right)_{ms} \left( \vec{\bf Q}_\sigma
\right)_s \nonumber \\
&\equiv& \sum_s {\cal M}_\sigma \left( {\cal O}^{(1)} {\cal O}^{(2)} \right)_{ns}
\left( {\vec {\bf Q}}_\sigma \right)_s \ ,
\label{EQ61} \end{eqnarray}
from which we might conclude that
\begin{eqnarray}
{\cal M}_\sigma ( {\cal O}^{(1)} {\cal O}^{(2)} ) &=&
{\cal M}_\sigma ( {\cal O}^{(2)}) {\cal M}_\sigma ({\cal O}^{(1)} ) \ .
\label{EQ62} \end{eqnarray}
One reason this result is wrong is that the first equation of Eq. (\ref{EQ61}) interprets
${\cal O}^{(1)} {\cal O}^{(2)} (\vec{\bf Q}_\sigma )_n$ to mean
${\cal O}^{(1)} [ {\cal O}^{(2)} ( \vec{\bf Q}_\sigma )_n ]$,
whereas Eq. (\ref{EQ62}) interprets it to mean $[ {\cal O}^{(1)} {\cal O}^{(2)} ]
( \vec{\bf Q}_\sigma )_n$. Another problem is that up to now, the operators 
${\cal O}^{(n)}$ operate on order parameters and {\it not} on  each other. This situation
is discussed in detail by Wigner.\cite{EPW} Instead we assert that
\begin{eqnarray}
{\cal M}_\sigma ( {\cal O}^{(1)}{\cal O}^{(2)} ) = 
{\cal M}_\sigma ( {\cal O}^{(1)}) {\cal M}_\sigma ( {\cal O}^{(2)} ) \ .
\label{EQ63} \end{eqnarray}
As an example of Eq. (\ref{EQ63}) consider the relation from Eq. (1) that
$2_z=m_ym_x \not= m_x m_y$.  Then, according to Eq. (\ref{EQ63}) we should have
\begin{eqnarray}
{\cal M}_\sigma (2_z) &=& {\cal M}_\sigma (m_y) {\cal M}_\sigma(m_x)
\not= {\cal M}_\sigma (m_x) {\cal M}_\sigma (m_y) \ ,
\end{eqnarray}
which the reader can verify using Eqs. (\ref{LHSEQ}), (\ref{EQYMAT}), and (\ref{TRANSX}).

\section{LANDAU FREE ENERGY}

\subsection{MINIMAL (UNCOPUPLED) MODEL FOR ORDER PARAMETERS}

We start by describing the symmetry of the model when the order parameters
$X_1$ and $X_3$ at zero wave vector and ${\bf Q}_{\bf X}^{(\sigma)}$ at wave vector
${\bf k}_X$ are not coupled to one another.  We can imagine that ordering
has developed via consecutive continuous transitions, as might happen for a suitable
set of parameters having the same symmetry as MGO, but quite different in detail.
Although this is not the experimental scenario, it will provide a correct description 
of the symmetries of the phase.  Thus we imagine $X_1$ and $X_2$ to be governed by
a free energy
\begin{eqnarray}
F_1,3&=& a_1 (T-T_1)X_1^2 + u_1 X_1^4 + a_3 (T-T_3)X_3^2 + u_3 X_3^4 \ ,
\end{eqnarray}
and similarly the incommensurate order parameters to be governed by a free
energy, the simplest form of which is
\begin{eqnarray}
F_X &=& a_{X,1} (T-T_{X1}) (|{\bf Q}_A^{(1)}|^2 + |{\bf Q}_B^{(1)}|^2)
+ a_{X,2} (T-T_{X2}) (|{\bf Q}_A^{(2)}|^2 + |{\bf Q}_B^{(2)}|^2)
+ {\cal O}(|{\bf Q}|^4 ) \ .
\end{eqnarray}
We point out that within such a simple theory
and barring an unphysical accidental degeneracy, ${\bf Q}_X^{(\sigma=1)}$
and ${\bf Q}_X^{(\sigma=2)}$ would not have the {\it same} wave vector
because the exchange interactions are never exactly isotropic in an
orthorhombic crystal.

However, if the equilibrium value of the two wave vectors are almost equal in a
simple approximation, then there are terms in the Landau free energy which lock the two
wave vectors into equality,\cite{JPCM} and we assume this to be the case here.  Then the
nature of the ordered phase is dictated by the form of the quartic and higher order
terms of the Landau free energy.  Consider, for example, the quartic terms.  In the space
of ${\bf Q}_A$ and ${\bf Q}_B$ there are isotropic terms
\begin{eqnarray}
\Delta F &=& \sum_\sigma u_\sigma \left[ |{\bf Q}_A^{(\sigma)}|^2
+ |{\bf Q}_B^{(\sigma)}|^2 \right]^2 \ .
\end{eqnarray}
This term would allow for an arbitrary superposition of both wave vectors, ${\bf k}_A$
and ${\bf k}_B$ within a single domain.  However, it has been shown\cite{HONDA}
that each domain contains only a single wave vector.  That indicates that the free
energy includes the term
\begin{eqnarray}
\Delta F &=& \sum_{\sigma \sigma'} B_{\sigma \sigma'}
|{\bf Q}_A^{(\sigma)}|^2 |{\bf Q}_B^{(\sigma')}|^2
\end{eqnarray}
which strongly disfavors having two wave vectors simultaneously present when 
$B_{\sigma \sigma'}$ is large and positive.  Finally, we point out that we
expect terms in the free energy to prevent irreps from having the same phase.
At positions where one irrep is maximal, there is usually less phase space into
which the other irrep can condense.  This argument is reflected by the term\cite{JPCM}
\begin{eqnarray}
\Delta F &=& A \sum_{X+A,B} \left( {\bf Q}_X^{(1)}[ {\bf Q}_X^{(2)}]^* +
[{\bf Q}_X^{(1)}]^* {\bf Q}_X^{(2)} \right)^2  \ ,
\label{EQ69} \end{eqnarray}
with $A>0$.  This term is proportional to $\cos^2 (\Delta \phi)$, where $\Delta \phi$ is 
the phase difference between the complex-valued order parameters of the two irreps.
We expect $A$ to be large and positive, in which case $\Delta \phi = \pm \pi /2$ 
is strongly favored.  The effects we attribute here to quartic terms could equally
well be attributed to higher order terms of similar symmetry.

\subsection{COUPLING TERMS IN THE FREE ENERGY}

\begin{table}
\caption{\label{SYMBB} SYMMETRY OF THE BUILDING BLOCKS}
\vspace*{0.2 in}
\begin{tabular} {|| c c | c || }
\hline \hline
& Operator &\ \ Transforms like\ \ \\
\hline
\ $F_1=$\ & $X_1X_3$ & $m_x m_y$ \\
\ $F_2=$ & $i \left[ {\bf Q}_A^{(1)} {{\bf Q}_A^{(2)}}^*
- {{\bf Q}_A^{(1)}}^* {{\bf Q}_A^{(2)}}
- {\bf Q}_B^{(1)} {{\bf Q}_B^{(2)}}^*
+ {{\bf Q}_B^{(1)}}^* {{\bf Q}_B^{(2)}}\right]$ \ \
& $m_z$ \\
\ $F_3=$ & $i \left[ {\bf Q}_A^{(1)} {{\bf Q}_A^{(2)}}^*
- {{\bf Q}_A^{(1)}}^* {{\bf Q}_A^{(2)}}
+ {\bf Q}_B^{(1)} {{\bf Q}_B^{(2)}}^*
- {{\bf Q}_B^{(1)}}^* {{\bf Q}_B^{(2)}}\right]$ \ \
& $m_xm_ym_z$ \\
\ $F_4=$ & $|{\bf Q}_A^{(1)}|^2 - |{\bf Q}_B^{(1)}|^2$ & $m_xm_y$ \\
\ $F_5=$ & $|{\bf Q}_A^{(2)}|^2 - |{\bf Q}_B^{(2)}|^2$ & $m_xm_y$ \\
\ $F_6=$ & $P_z$ & $m_z$ \\
\hline \hline
\end{tabular}
\end{table}

Before proceeding to higher order we emphasize that we only want to
enumerate the lowest order terms which have each possible allowed symmetry.
To construct such higher than quadratic order terms which are allowed by
symmetry, we formulate the following rules. Rule 1: we do not allow a
term which includes a factor which itself transforms like unity, such as
$X_n^2$ or $|{\bf Q}_X^{(\sigma)}|^2$, because the term without this 
factor should already be in our list of allowed terms. Such terms do not
lead to a different symmetry.  They only make a quantitative change
in the response of the system.  Rule 2: Due to
translational invariance, an allowed term must conserve wave vector.  
In view of Rule 1, this rule implies that the incommensurate order 
parameters can only occur in the combination 
${\bf Q}_X^{(\sigma)}[{\bf Q}_X^{(\sigma')}]^*$.\cite{UMKL}
Rule 3: due to time reversal invariance any term must contain an even
number of magnetic order parameters.  In view of the previous rules,
$X_1$ and $X_3$ can only occur in the product
$X_1X_3$. To summarize: the allowed building blocks for invariants
are a) $X_1X_3$,
b) $i \left( {\bf Q}_X^{(1)} [{\bf Q}_X^{(2)}]^*
 - [{\bf Q}_X^{(1)}]^* {\bf Q}_X^{(2)} \right)$, 
c) $|{\bf Q}_A^{(\sigma)}|^2 - |{\bf Q}_B^{(\sigma)}|^2$,
d) components of the electric polarization, ${\bf P}$ or 
magnetization ${\bf M}$.
Similar terms in which ${\bf P}$ is replaced by the applied electric
field or ${\bf M}$ is replaced by the applied magnetic field are also
possible, but are not considered here. Note: the term
$\left( {\bf Q}_X^{(1)}[ {\bf Q}_X^{(2)}]^* + [{\bf Q}_X^{(1)}]^* {\bf Q}_X^{(2)}\right)$
is excluded by the potential of Eq. (\ref{EQ69}).
In Table \ref{SYMBB} we list the symmetry of the various building blocks. 
To verify these results use Eqs. (\ref{TRANSZ}), (\ref{EQYMAT}), and (\ref{TRANSX}) to write
\begin{eqnarray}
m_x {\bf Q}_A^{(1)} {{\bf Q}_A^{(2)}}^* &=& {{\bf Q}_B^{(1)}}^* {\bf Q}_B^{(2)} \ , \ \ \ \
m_x {\bf Q}_B^{(1)} {{\bf Q}_B^{(2)}}^* = {{\bf Q}_A^{(1)}}^* {\bf Q}_A^{(2)} \ ,
\nonumber \\
m_y {\bf Q}_A^{(1)} {{\bf Q}_A^{(2)}}^* &=& - {\bf Q}_B^{(1)} {{\bf Q}_B^{(2)}}^* \ , \ \ \ \
m_y {\bf Q}_B^{(1)} {{\bf Q}_B^{(2)}}^* = - {\bf Q}_A^{(1)} {{\bf Q}_A^{(2)}}^* \ ,
\nonumber \\
m_z {\bf Q}_A^{(1)} {{\bf Q}_A^{(2)}}^* &=& - {\bf Q}_A^{(1)} {{\bf Q}_A^{(2)}}^* \ , \ \ \ \
m_z {\bf Q}_B^{(1)} {{\bf Q}_B^{(2)}}^* = - {\bf Q}_B^{(1)} {{\bf Q}_B^{(2)}}^* \ .
\label{EQ70} \end{eqnarray}
One can check that, in agreement with Table VI,
\begin{eqnarray}
m_x F_2 = m_y F_2 =- m_z F_2 = F_2 \ , \ \ \ \ \ \
m_x F_3 = m_y F_3 = m_z F_3 = -F_3 \ .
\end{eqnarray}

The simplest invariant is
\begin{eqnarray}
U&=& aF_2F_6 =  i a \left[ {\bf Q}_A^{(1)} {{\bf Q}_A^{(2)}}^*
- {{\bf Q}_A^{(1)}}^* {{\bf Q}_A^{(2)}}
- {\bf Q}_B^{(1)} {{\bf Q}_B^{(2)}}^*
+ {{\bf Q}_B^{(1)}}^* {{\bf Q}_B^{(2)}}\right] P_z \ ,
\end{eqnarray}
which is the usual trilinear magnetoelectric interaction which
induces a nonzero equilibrium value of $P_z$.\cite{GL,MK}
Then we have the invariants
\begin{eqnarray}
V_1 &=& b_1F_1F_4 = b_1X_1X_3 \left[ 
|{\bf Q}_A^{(1)}|^2 - |{\bf Q}_B^{(1)}|^2\right]  \\
V_2&=& b_2 F_1 F_5 = b_2 X_1 X_3 \left[ |{\bf Q}_A^{(2)}|^2 - |{\bf Q}_B^{(2)}|^2\right] \ . 
\end{eqnarray}
In principle we could also list $F_4F_5$.  But when each domain only has a
single wave vector, this term is not interesting.
Finally we have $W \equiv F_1F_3F_6$:
\begin{eqnarray}
W &=& i c X_1X_3P_z \left[ {\bf Q}_A^{(1)} {{\bf Q}_A^{(2)}}^*
- {{\bf Q}_A^{(1)}}^* {{\bf Q}_A^{(2)}}
+ {\bf Q}_B^{(1)} {{\bf Q}_B^{(2)}}^*
- {{\bf Q}_B^{(1)}}^* {{\bf Q}_B^{(2)}}\right] \ .
\end{eqnarray}
Again, we omit the terms $F_3F_4F_6$ and $F_3F_5F_6$ because when each domain
only has a single wave vector, these interactions are not interesting.
The consequences of the potentials $U$, $V$, and $W$ for the switching
behavior of MGO are discussed in detail in Ref. \onlinecite{HONDA}.

We should also note the existence of the invariant
\begin{eqnarray}
Y &=& \left[ e X_1X_3 + \sum_\sigma f_\sigma \left(
|{\bf Q}_A^{(\sigma)}|^2 - |{\bf Q}_B^{(\sigma)}|^2 \right) \right]
P_x P_y \ .
\end{eqnarray}
This term shows that in this phase the electric susceptibility tensor
has off-diagonal $x$-$y$ elements induced by the magnetic ordering
which depend on the wave vector of the domain and on $X_3$, the
$z$-component of magnetization.

\subsection{EQUILIBRIUM PHASES}

In this section we minimize the free energy including coupling terms and thereby 
determine the various equilibrium domains that are possible.  This discussion 
is {\it not} equivalent to discussing the switching between equilibrium states.
To illustrate the difference between these two analyses consider the following
two scenarios.  In scenario I, one simply cools into the lowest temperature
phase and then asks if there is any correlation between the orientation of the
net magnetization and that of the net ferroelectric polarization: i. e. in any 
arbitrarily selected domain are these two collinear vectors always parallel or always
antiparallel to one another? The experimental answer is ``no:''\cite{MGO1} in some
domains the two vectors are parallel and in other domains they are antiparallel.
In scenario II, one asks a different question: if the magnetic field is used to
reverse the direction of the magnetization in a domain, will that always cause the
direction of the ferroelectric polarization in that domain to reverse? The
experimental answer to that question is ``yes.''\cite{HONDA}

Here we only consider the equilibrium properties and we rely on the experimental
observation that each domain contains only one of the two possible wave vectors.\cite{HONDA}
So we have therefore two choices for domains: they have either wave vector 
$\pm {\bf k}_A$ or  wave vector $\pm {\bf k}_B$.  We assume that the magnitudes 
of the order parameters ${\bf Q}_X^{(\sigma)}$ are fixed by the terms in the free
energy which only depend on these variables, and domains A and B are related by
$|{\bf Q}_A^{(\sigma)}|  = |{\bf Q}_B^{(\sigma)}|$.  We assume that the 
{\it magnitudes} of ${\bf X}_1$ and ${\bf X}_3$ (but not their algebraic signs) 
are similarly fixed.  When we minimize the $U$, $V$, and $W$ terms it is
obvious that we will only be able to determine the product of $X_1$ and $X_3$.  
Accordingly, we will have domains in which the sign and magnitude of $X_1X_3$
is determined so as to minimize the free energy, but the algebraic sign of $X_3$,
the magnetization, can be chosen arbitrarily while keeping the sign of the product
$X_1X_3$ fixed.  Finally,
since the free energy clearly does not depend on the sign of the ferroelectric
polarization, $P_z$, we will have domains with either sign of $P_z$.  So in all,
these three independent binary choices gives rise to eight possible distinct
domains, as already noted in Ref. [\onlinecite{HONDA}].

Accordingly, for a domain of wave vector ${\bf k}_X$ we introduce the variables
$\eta_X$ and $\zeta_X$ by
\begin{eqnarray}
\eta_X & \equiv& X_1X_3/|X_1X_3| \ , \ \ \ \ \ \
{\bf Q}_X^{(1)} {{\bf Q}_X^{(2)}}^* - {{\bf Q}_X^{(1)}}^* {\bf Q}_X^{(2)} =
2 i \zeta_X | {\bf Q}_X^{(1)} {\bf Q}_X^{(2)}  |
\end{eqnarray}
so that in the domain of ${\bf k}_X$
$\eta_X = \pm 1$ is the phase of the staggered magnetization relative to
that of the ferromagnetic moment
and $\zeta_X= \pm 1$ is the phase of ${\bf Q}_X^{(1)}$
relative that of ${\bf Q}_X^{(2)}$ in units of $\pi/2$ (which is the
definition of the helicity).  We first consider the situation in the 
domain of wave vector ${\bf k}_A$.  There we have the free energy
\begin{eqnarray}
F_A &\equiv& U + V_1 + V_2 + W = -2a P_z \zeta_A | {\bf Q}_A^{(1)} {\bf Q}_A^{(2)}|
+ \eta_A \left[ b_1 | {\bf Q}_A^{(1)}|^2 +  b_2 | {\bf Q}_A^{(2)}|^2 \right]
|X_1 X_3| \nonumber \\
&& -2 c \eta_A P_z \zeta_A |X_1 X_3| |{\bf Q}_A^{(1)} {\bf Q}_A^{(2)} | + \frac{1}
{2 \chi_E} P_z^2 \ ,
\end{eqnarray}
where we have now included the free energy due to the polarization in terms of the
electric polarizability $\chi_E$.  We minimize the free
energy $F_A$ with respect to $P_z$, so that
\begin{eqnarray}
\chi_E^{-1} P_z &=& 2 \zeta_A \left(  a + c \eta_A |X_1 X_3| \right) |{\bf Q}_A^{(1)}
{\bf Q}_A^{(2)} | \ ,
\label{EQ80} \end{eqnarray}
in which case the equilibrium free energy becomes (since $\zeta_A^2 =1$)
\begin{eqnarray}
F_A &=& -2 \chi_E \left( a + c \eta_A |X_1 X_3| \right)^2
|{\bf Q}_A^{(1)} {\bf Q}_A^{(2)}|^2 \nonumber \\
&& + \eta_A \left[ b_1 |{\bf Q}_A^{(1)}|^2 + b_2 |{\bf Q}_A^{(2)}|^2 \right]
|X_1 X_3 | \ .
\end{eqnarray}
As expected, $F_A$ does not depend on the sign of the helicity $\zeta_A$ which
determines the sign of $P_z$.  We need to minimize this with respect to $\eta_A$. 
We write
\begin{eqnarray}
F_A &=& -2 \chi_E \left( a^2 + c^2 |X_1X_3|^2 \right) | {\bf Q}_A^{(1)} {\bf Q}_A^{(2)}|^2
\nonumber \\ && + \eta_A |X_1X_3| \left( b_1 |{\bf Q}_A^{(1)}|^2
+ b_2|{\bf Q}_A^{(2)}|^2 -4 \chi_E ca |{\bf Q}_A^{(1)} {\bf Q}_A^{(2)}|^2 \right) \ , 
\end{eqnarray}
so that $\eta_A = - {\cal R}_A /|{\cal R}_A|$, where
\begin{eqnarray}
{\cal R}_X &=& b_1 | {\bf Q}_X^{(1)}|^2 + b_2 |{\bf Q}_X^{(2)}|^2
-4 \chi_E ca | {\bf Q}_X^{(1)} {\bf Q}_X^{(2)} |^2 \  ,
\end{eqnarray}
where $X$ is either $A$ or $B$. Thus in a domain of wave vector ${\bf k}_A$
\begin{eqnarray}
\chi_E^{-1} P_z &=& 2 \zeta_A \left( a - c {\cal R}_A
|X_1X_3/{\cal R}_A| \right) | {\bf Q}_A^{(1)}  {\bf Q}_A^{(2)}| \ .
\end{eqnarray}
The analysis for wave vector ${\bf k}_B$ is similar and yields the result
\begin{eqnarray}
\chi_E^{-1} P_z &=& - 2 \zeta_A \left( a - c {\cal R}_B
|X_1X_3/{\cal R}_B| \right) | {\bf Q}_B^{(1)}  {\bf Q}_B^{(2)}| \ .
\end{eqnarray}
All domains have the same magnitude of $P_z$, but its sign
varies from domain to domain.

To summarize: each domain is characterized by 1) the axis of the wave vector,
${\hat {\bf k}}_A$ or ${\hat {\bf k}}_B$,
2) the sign of $P_z$ [or equivalently, according to Eq. (\ref{EQ80}),
the sign of the helicity $\zeta_X$], and 3) the sign of the net
magnetization along $\hat c$.  But all domains are symmetry-related
to one another.

\section{CONCLUSIONS}

Many of our results for the system Mn$_2$GeO$_4$ (MGO) have useful analogs for
other noncollinear magnetic incommensurate systems.  For instance,
note the way we simplified the output of ISODISTORT in the Appendix A.
Secondly, the case of MGO illustrates how one introduces order
parameters as the amplitudes of the magnetic ``modes.''  This formulation
is reminiscent of the description of lattice vibration in terms
of normal modes amplitudes and the symmetry analysis that follows here
parallels that of phonon modes.  A significant advantage of introducing
order parameters is that they conveniently carry with them the 
symmetry properties of the modes.  In the usual scenario involving
magnetic order parameters, the symmetry is trivial.  Here in a more
complex setting, the analysis is more involved and one has to keeep track
of what is called here the ``wave function.''  The wave function gives
meaning to the order parameters. This end result is well 
illustrated by the application of the results of this paper to the
switching phenomena described in Ref. \onlinecite{HONDA}.

\vspace{0.2 in}
{\bf ACKNOWLEDGEMENT}

I thank M. Kenzelmann for inviting me to collaborate in
the study of Mn$_2$GeO$_4$ and for many clarifying discussions
of the results of Ref. \onlinecite{HONDA}. I also
acknowledge discussions concerning irreps  with J. White and
using ISODISTORT with B. Campbell. I thank the Department of Commerce
for supporting work done at the National Institute of Standards and
Technology.
\begin{appendix}
\section{Modes for $k_A$}

\subsection{Modes for Irreps D$^{(1)}$ and D$^{(2)}$}

In Tables \ref{MODE1} and \ref{MODE2} we show the modes for irrep $D^{(1)}$
for ${\bf k}_A = (k_x,k_y,0)$.  Table \ref{MODE3} and \ref{MODE4}
give the analogous data for irrep $D^{(2)}$. From these tables (taken from 
ISODISTORT) we obtain the magnetization distribution throughout a domain
for irrep $\sigma$ as
\begin{eqnarray}
{\bf M}({\bf N} + \tauv_n) = {\bf m}^{(\sigma)} (\tauv_n) 
e^{i \chi -i {\bf k} \cdot ( {\bf N} + \tauv_n)} + {\rm c. \ c.} \ .
\end{eqnarray}
We allow the modes to have an arbitrary overall phase factor, $\exp(i \chi)$, because
the origin of the incommensurate excitation is arbitrary.  We now write the results
of ISODISTORT given in Tables I and II in a simpler, but equivalent form.  For irrep
D$^{(1)}$ for the ch sites we make the cosmetic replacement
\begin{eqnarray}
m_\alpha^{(1)}(\tauv_1)&=& a_\alpha \ , \ \ \
m_\alpha^{(1)}(\tauv_2)= \mu_\alpha a_\alpha \ , \ \ \
m_\alpha^{(1)}(\tauv_3)= b_\alpha \ , \ \ \
m_\alpha^{(1)}(\tauv_4)= \mu_\alpha b_\alpha \ ,
\end{eqnarray}
where $(\mu_x,\mu_y, \mu_z)=(-1,-1,+1)$.  For the pl sites, the sitation is more
complicated. Table VIII gives the terms from ISODISTORT which depend on $Z_3$ and
$Z_4$ as
\begin{eqnarray}
m_x^{(1)}(\tauv_5) &=& (Z_3-iZ_4) e^{i \phi} \ , \ \ \ 
m_x^{(1)}(\tauv_6) = -(Z_3-iZ_4) e^{i \phi} \ , \nonumber \\
m_x^{(1)}(\tauv_7) &=& (Z_3+iZ_4) e^{-i \phi} \ , \ \ \ 
m_x^{(1)}(\tauv_8) = -(Z_3+iZ_4) e^{-i \phi} \ .
\end{eqnarray}

\begin{table} [h!]
\caption{\label{MODE1} 
Structure of the six modes for ${\bf k} = {\bf k}_A \equiv (0.138,0.211,0)$,\cite{ISO}
irrep $D^{(1)}$, for ch sites  $\tauv_1$ - $\tauv_4$ from ISODISTORT
for space group Pnma = No. 62 in Ref. \onlinecite{ITC}.
The magnetic moment distribution is ${\bf M}({\bf R} + \tauv_n)=
e^{- i {\bf k} \cdot ({\bf R}+ \tauv_n)}{\bf m}(\tauv_n)$, where
the first column lists the real-valued amplitudes which give
${\bf m}(\tauv_n)$, where the ${\bf m}(\tauv_n)$ are  listed in the form
$(R,\theta) \equiv R\exp (i \theta)$.  An arbitrary overall phase, the
same for ch and pl sites, is not included in these tables.}
\vspace{0.2 in}
\begin{tabular} {|| c | l | c c c ||  l | c c c ||}
\hline
AMP &  &\ \ $m_x$ & $m_y$ & $m_z$  &
&\ \ $m_x$ & $m_y$ & $m_z$ \\
\hline
$Y_1$ & $\tauv_1$ & (1,\ 0)    & (0,\ 0)   & (0,\ 0) 
      & $\tauv_2$ & (1,\ 180)  & (0,\ 0)  & (0,\ 0) \\
\hline
$Y_2$ & $\tauv_3$ & (1,\ 0)    & (0,\ 0)   & (0,\ 0) 
      & $\tauv_4$ & (1,\ 180)  & (0,\ 0)   & (0,\ 0) \\
\hline
$Y_3$ & $\tauv_1$ & (0,\ 0)    & (1,\ 0)   & (0,\ 0)  
      & $\tauv_2$ & (0,\ 0)    & (1,\ 180) & (0,\ 0) \\
\hline
$Y_4$ & $\tauv_3$ & (0,\ 0)    & (1,\ 180)$^{\rm a}$ & (0,\ 0)
      & $\tauv_4$ & (0,\ 0)    & (1,\ 0)$^{\rm a}$ & (0,\ 0) \\
\hline
$Y_5$ & $\tauv_1$ & (0,\ 0)    & (0,\ 0)   & (1,\ 0)  
      & $\tauv_2$ & (0,\ 0)    & (0,\ 0)   & (1, \ 0) \\
\hline
$Y_6$ & $\tauv_3$ & (0,\ 0)    & (0,\ 0)   & (1,\ 180)$^{\rm a}$
      & $\tauv_4$ & (0,\ 0)    & (0,\ 0)   & (1,\ 180)$^{\rm a}$ \\
\hline \hline
\end{tabular}

\vspace*{0.15 in}
\noindent
\noindent
a) We will reparametrize: $Y_6 \rightarrow -Y_6$ and
$Y_4 \rightarrow -Y_4$, so that these
phases are regularized.
\end{table}

\noindent
Note that this is a parametrization in terms of three parameters.  However, this is an
overparametrization: if $\phi$ is arbitrarily varied, $m_x$ can remain unchanged by a suitable
rotation in the complex $(Z_3,Z_4)$ plane.  Accordingly, we reproduce these result via a 
two-parameter representation in terms of the complex-valued variable
$z_x = (Z_3-iZ_4)e^{i \phi}$, so that
\begin{eqnarray}
m_x^{(1)}(\tauv_5) &=& z_x \ , \ \ \
m_x^{(1)}(\tauv_6) = -z_x \ , \ \ \
m_x^{(1)}(\tauv_7) = z_x^* \ , \ \ \
m_x^{(1)}(\tauv_8) = -z_x^* \ .
\end{eqnarray}
Similarly, we can reproduce the results of Table \ref{MODE2} from ISODISTORT
for $m_y$ on the pl sites by setting $z_y=(Z_1-iZ_2)e^{i \phi}$ in which case
\begin{eqnarray}
m_y^{(1)}(\tauv_5) &=& (Z_1 -iZ_2)e^{i \phi} = z_y \ , \ \ \
m_y^{(1)}(\tauv_6) = - (Z_1 -iZ_2)e^{i \phi} = - z_y \ , \nonumber \\
m_y^{(1)}(\tauv_7) &=& (Z_1 +iZ_2)e^{-i \phi} = z_y^* \ , \ \ \
m_y^{(1)}(\tauv_8) = -(Z_1 + iZ_2)e^{-i \phi} = -z_y^* \ , \ \ \
\end{eqnarray}
and for $M_z$ on the pl sites by setting $z_z=(Z_5-iZ_6)e^{i \phi}$ in which case
\begin{eqnarray}
m_z^{(1)}(\tauv_5) &=& (Z_5 -iZ_6)e^{i \phi} = z_z \ , \ \ \
m_z^{(1)}(\tauv_6) =  (Z_5 -iZ_6)e^{i \phi} = z_z \ , \nonumber \\
m_z^{(1)}(\tauv_7) &=& (Z_5 +iZ_6)e^{-i \phi} = z_z^* \ , \ \ \
m_z^{(1)}(\tauv_8) = (Z_5 + iZ_6)e^{-i \phi} = z_z^* \ , \ \ \
\end{eqnarray}
Similar identifications are mode for irrep D$^{(2)}$ and we obtain Eqs. 
(\ref{NOTE}) and (\ref{MODEQ2}).

\begin{table} [h!]
\caption{\label{MODE2}
As Table VII.  Mode structure for ${\bf k} = (0.138,0.211,0)$,
irrep D$^{(1)}$, for pl sites.  ISODISTORT sets $\phi=327.55$, but
as we discuss, this value has no significance.}
\vspace{0.3 in}
\begin{tabular} {|| c | c | c c c || c | c c c ||}
\hline
AMP &  &\ \ $m_x$ & $m_y$ & $m_z$ & \ \ & $m_x$ & $m_y$ & $m_z$ \\
\hline
$Z_1$ & $\tauv_5$  & (0,\ 0)           & (1,\ $\phi$)      & (0,\ 0)
      & $\tauv_6$  & (0,\ 0)           & (1,\ $\phi -180)$ & (0,\ 0) \\
      & $\tauv_7$  & (0,\ 0)           & (1,\ $360-\phi )$ & (0,\ 0)
      & $\tauv_8$  & (0,\ 0)           & (1,\ $540-\phi )$ & (0,\ 0) \\
\hline \hline
$Z_2$ & $\tauv_5$  & (0,\ 0)           & (1,\ $\phi-90)$   & (0,\ 0) 
      & $\tauv_6$  & (0,\ 0)           & (1,\ $\phi-270)$  & (0,\ 0) \\
      & $\tauv_7$  & (0,\ 0)           & (1,\ $450-\phi)$  & (0,\ 0)
      & $\tauv_8$  & (0,\ 0)           & (1,\ $630-\phi)$  & (0,\ 0) \\
\hline \hline
$Z_3$ & $\tauv_5$  & (1,\ $\phi$)      & (0,\ 0)           & (0,\ 0) 
      & $\tauv_6$  & (1,\ $\phi-180)$  & (0,\ 0)           & (0,\ 0) \\
      & $\tauv_7$  & (1,\ $360-\phi)$  & (0,\ 0)           & (0,\ 0) 
      & $\tauv_8$  & (1,\ $540-\phi)$  & (0,\ 0)           & (0,\ 0) \\
\hline \hline
$Z_4$ & $\tauv_5$  & (1,\ $\phi-90)$   & (0,\ 0)           & (0,\ 0) 
      & $\tauv_6$  & (1,\ $\phi-270)$  & (0,\ 0)           & (0,\ 0) \\
      & $\tauv_7$  & (1,\ $450-\phi)$  & (0,\ 0)           & (0,\ 0) 
      & $\tauv_8$  & (1,\ $630-\phi)$  & (0,\ 0)           & (0,\ 0) \\
\hline \hline
$Z_5$ & $\tauv_5$  & (0,\ 0) & (0,\ 0) & (1,\ $\phi$) 
      & $\tauv_6$  & (0,\ 0) & (0,\ 0) & (1,\ $\phi)$ \\
      & $\tauv_7$  & (0,\ 0) & (0,\ 0) & (1,\ $360-\phi)$ 
      & $\tauv_8$  & (0,\ 0) & (0,\ 0) & (1,\ $360-\phi)$ \\
\hline \hline
$Z_6$ & $\tauv_5$  & (0,\ 0) & (0,\ 0) & (1,\ $\phi-90)$ 
      & $\tauv_6$  & (0,\ 0) & (0,\ 0) & (1,\ $\phi-90)$ \\
      & $\tauv_7$  & (0,\ 0) & (0,\ 0) & (1,\ $450-\phi)$ 
      & $\tauv_8$  & (0,\ 0) & (0,\ 0) & (1,\ $450-\phi)$ \\
\hline \hline
\end{tabular}
\end{table}

\begin{table} [h!]
\caption{\label{MODE3}
As Table VII. Mode structure for ${\bf k}=(0.138,0.211,0)$ for
irrep $D^{(2)}$}
\vspace{0.3 in}
\begin{tabular} {|| c | l | c c c ||  l | c c c ||}
\hline
AMP &  &\ \ $m_x$ & $m_y$ & $m_z$ &\ \ & $m_x$ & $m_y$ & $m_z$ \\
\hline
$Y_1$ & $\tauv_1$ & (1,\ 270)  & (0,\ 0)   & (0,\ 0) 
      & $\tauv_2$ & (1,\ 270)  & (0,\ 0)   & (0,\ 0) \\
\hline \hline
$Y_2$ & $\tauv_3$ & (1,\ 270)  & (0,\ 0)   & (0,\ 0) 
      & $\tauv_4$ & (1,\ 270)  & (0,\ 0)   & (0,\ 0) \\
\hline \hline
$Y_3$ & $\tauv_1$ & (0,\ 0)    & (1,\ 270) & (0,\ 0) 
      & $\tauv_2$ & (0,\ 0)    & (1,\ 270) & (0,\ 0) \\
\hline \hline
$Y_4$ & $\tauv_3$ & (0,\ 0)    & (1,\ 90)  & (0,\ 0)
      & $\tauv_4$ & (0,\ 0)    & (1,\ 90)  & (0,\ 0) \\
\hline \hline
$Y_5$ & $\tauv_1$ & (0,\ 0)    & (0,\ 0)   & (1,\ 270) 
      & $\tauv_2$ & (0,\ 0)    & (0,\ 0)   & (1, \ 90) \\
\hline \hline
$Y_6$ & $\tauv_3$ & (0,\ 0)    & (0,\ 0)   & (1,\  90)
      & $\tauv_4$ & (0,\ 0)    & (0,\ 0)   & (1,\ 270) \\
\hline \hline
\end{tabular}
\end{table}

\begin{table} [h!]
\caption{\label{MODE4}
As Table VII.  Mode structure for ${\bf k} = (0.138,0.211,0)$
for irrep D$^{(2)}$.}
\vspace{0.3 in}
\begin{tabular} {|| c | c | c c c || c | c c c ||}
\hline
AMP &  &\ \ $m_x$ & $m_y$ & $m_z$ & & \ \ $m_x$ & $m_y$ & $m_z$ \\
\hline
$Z_1$ & $\tauv_5$  & (0,\ 0)           & (1,\ $\phi$)      & (0,\ 0)
      & $\tauv_6$  & (0,\ 0)           & (1,\ $\phi )$     & (0,\ 0) \\
      & $\tauv_7$  & (0,\ 0)           & (1,\ $540-\phi )$ & (0,\ 0)
      & $\tauv_8$  & (0,\ 0)           & (1,\ $540-\phi )$ & (0,\ 0) \\
\hline \hline
$Z_2$ & $\tauv_5$  & (0,\ 0)           & (1,\ $\phi-90)$   & (0,\ 0) 
      & $\tauv_6$  & (0,\ 0)           & (1,\ $\phi-90)$   & (0,\ 0) \\
      & $\tauv_7$  & (0,\ 0)           & (1,\ $630-\phi)$  & (0,\ 0)
      & $\tauv_8$  & (0,\ 0)           & (1,\ $630-\phi)$  & (0,\ 0) \\
\hline \hline
$Z_3$ & $\tauv_5$  & (1,\ $\phi$)      & (0,\ 0)           & (0,\ 0) 
      & $\tauv_6$  & (1,\ $\phi)$      & (0,\ 0)           & (0,\ 0) \\
      & $\tauv_7$  & (1,\ $540-\phi)$  & (0,\ 0)           & (0,\ 0) 
      & $\tauv_8$  & (1,\ $540-\phi)$  & (0,\ 0)           & (0,\ 0) \\
\hline \hline
$Z_4$ & $\tauv_5$  & (1,\ $\phi-90)$   & (0,\ 0)           & (0,\ 0) 
      & $\tauv_6$  & (1,\ $\phi-90)$   & (0,\ 0)           & (0,\ 0) \\
      & $\tauv_7$  & (1,\ $630-\phi)$  & (0,\ 0)           & (0,\ 0) 
      & $\tauv_8$  & (1,\ $630-\phi)$  & (0,\ 0)           & (0,\ 0) \\
\hline \hline
$Z_5$ & $\tauv_5$  & (0,\ 0) & (0,\ 0) & (1,\ $\phi$) 
      & $\tauv_6$  & (0,\ 0) & (0,\ 0) & (1,\ $\phi)-180$ \\
      & $\tauv_7$  & (0,\ 0) & (0,\ 0) & (1,\ $540-\phi)$ 
      & $\tauv_8$  & (0,\ 0) & (0,\ 0) & (1,\ $360-\phi)$ \\
\hline \hline
$Z_6$ & $\tauv_5$  & (0,\ 0) & (0,\ 0) & (1,\ $\phi-90)$ 
      & $\tauv_6$  & (0,\ 0) & (0,\ 0) & (1,\ $\phi-270)$ \\
      & $\tauv_7$  & (0,\ 0) & (0,\ 0) & (1,\ $630-\phi)$ 
      & $\tauv_8$  & (0,\ 0) & (0,\ 0) & (1,\ $450-\phi)$ \\
\hline \hline
\end{tabular}
\end{table}

\section{TEMPERATURE DEPENDENCE OF MODES}

Look at Eq. (\ref{ALLEQ}).  There one sees that each mode involves
12 real parameters.  Thus, there are actually 11 additional modes having
the same symmetry as the mode we focus upon.  Thus we introduce
corresponding mode amplitudes $Q_n$, with $n=1,12$, where the free
energy at quadratic order due to the irrep in question
in the disordered phase is
\begin{eqnarray}
F  &=& \sum_{n=1}^{12} F_n (T_n-T) |Q_n|^2 \ ,
\end{eqnarray}
where $T_1$ is the largest $T_n$, so that the mode labeled ``1''
is the one that first condenses as the temperature is lowered.
To study the mean-field temperature-dependence for $T$ just below
$T_1$ we go to higher order:
\begin{eqnarray}
F  &=& - a_1 (T_1-T) |Q_1|^2 + \sum_{n>1} (T_n-T) |Q_n^2|
+ u |Q_1|^4 + V \ ,
\end{eqnarray}
where $T_1<T$, $T_n > T$ for all $n>1$, and $u>0$.  Thereby we find
the standard result: $\langle Q_1 \rangle =0$, for $T>T_1$ and
for $T< T_1$
\begin{eqnarray}
|\langle Q_1 \rangle | = [ a_1/(2u)]^{1/2} [T_1-T]^{1/2} \ ,
\end{eqnarray}
where $\langle \ \ \rangle$ denotes an equilibrium  value.
In the present case, the terms which modify the critical
wave function associated with $Q_1$ are all even order
in the order parameters. The quadratic terms are diagonal
by construction of the normal modes.  So the leading term which
give corrections to the wave function of the critical mode is of the form
\begin{eqnarray}
V &=& \sum_{n>1} |Q_1|^2 \left[ c_n Q_1 Q_n^* + c_n^* Q_1^* Q_n \right]  \ ,
\end{eqnarray}
where $c_n$ need not be real-valued.
For $T<T_1$ the effect of this term is approximately the same as that of
\begin{eqnarray}
V &=& 2 \sum_{n>1} |\langle Q_1 \rangle|^2 \left[ c_n Q_1Q_n^*
+ c_n^* Q_1^* Q_n \right]   \ ,
\end{eqnarray}
Thus we see that at quartic order there
is a mixing of modes governed by the temperature dependent
prefactor proportional to the equilibrium
value,   $|\langle Q_1\rangle|^2$, which in mean field theory is proportional
to $T_1-T$. Of course, this mixing only takes place within the
space of modes having the same symmetry as $Q_1$.

\section{PHASE FACTORS}

In this section we discuss how the definition of order parameters is subject to
inclusion of arbitrary phase factors.  As mentioned, this ambiguity is similar to
that encountered in a two-sublattice antiferromagnet where one defines the
staggered magnetization order paramter ${\bf N}$ in terms of the sublattice
magnetizations, either as ${\bf N} = {\bf M}_1 - {\bf M}_2$ or as
${\bf N} = {\bf M}_2 - {\bf M}_1$.  A macroscopic observable will not 
depend on the sign of ${\bf N}$.  We now see how such a phase factor
affects our analysis.  Equations (\ref{EQPSIB1}) and (\ref{EQPSIB2}) define
the wave functions $\Psi_{\alpha n}^{(B,\sigma)}$.  In principle we can
introduce arbitrary phases $\exp(i \phi_\sigma)$ as prefactors in
these definitions.  It is not useful to go to that level of generality.  
So we will modify these definitions by writing
\begin{eqnarray}
\Psi^{(B,1)}_{\alpha n} &=& \xi_1 \lambda_\alpha' \Psi_{\alpha \overline n}^{(A,1)} =
\xi_1 \lambda'_\alpha [ \mu_\alpha b_\alpha,
b_\alpha , \mu_\alpha a_\alpha , a_\alpha ;
z_\alpha , \mu_\alpha z_\alpha , z_\alpha^* , \mu_\alpha z_\alpha^* ]_n \ ,  \\
\Psi^{(B,2)}_{\alpha n} &=& -\xi_2 \lambda_\alpha' \Psi_{\alpha \overline n}^{(A,2)}
= - \xi_2\lambda'_\alpha [ \mu_\alpha d_\alpha,
-d_\alpha , \mu_\alpha c_\alpha , -c_\alpha ;
-w_\alpha , \mu_\alpha w_\alpha , -w_\alpha^* , \mu_\alpha w_\alpha^* ]_n \ .
\end{eqnarray}
with each $\xi_\sigma = \pm 1$.  Then, in this version of the theory we have
\begin{eqnarray}
{\cal M}_\sigma (m_y) &=& (-1)^{\sigma+1} \left[ \begin{array} {c c c c}
0 & \xi_1 e^{ik_y/2} & 0 & 0 \\ \xi_2 e^{-ik_y/2}  & 0 & 0 & 0 \\
0 & 0 & 0 & \xi_1 e^{-ik_y/2} \\
0 & 0 & \xi_2 e^{ik_y/2} & 0 \\
\end{array} \right] \ . 
\end{eqnarray}
These modification do not affect ${\cal M}_\sigma (m_z)$ or
${\cal M}_\sigma (2_z)$ because $m_z$ and $2_z$ transform wave
functions into themselves.  Similarly, we now have the modified result
\begin{eqnarray}
{\cal M}_\sigma (m_x) &=&  (-1)^{\sigma+1} \left[ \begin{array} {c c c c}
0 & 0 & 0 & \xi_2 e^{i(k_x+k_y)/2} \\ 0 & 0 & \xi_1e^{i(k_x-k_y/2)}  & 0 \\
0 & \xi_2 e^{-i(k_x+k_y)/2} & 0 & 0 \\
\xi_1 e^{-i(k_x-k_y)/2} & 0 & 0 & 0 \\
\end{array} \right] \ . 
\end{eqnarray}
One can verify that these choices of phase to not affect Eq. (\ref{EQ70}) and
thus do not affect the results for the invariant potential $U$, $V_n$, and $W$.

\end{appendix}


\begin{thebibliography} {99}
\bibitem{UMKL} 
Umklapp terms would have to be of very high order and should be
completely negligible here.
\bibitem{EXACT} 
The exact values of the nonzero components of the wave vector clearly
do not influence the mode structure.  The value of $k_x=0.136$ differs
slightly from $k_x=0.138$ given in Ref. \onlinecite{MGO1}.
\end{thebibliography}
\end{document}